\documentclass[aps,floatfix,nofootinbib,amsmath,amssymb,amsfonts,notitlepage,superscriptaddress, twocolumn]{revtex4-2}

\usepackage[T1]{fontenc}
\usepackage[latin9]{inputenc}
\usepackage{lmodern} 
\usepackage{color}
\usepackage{bbold}
\usepackage{dsfont}
\usepackage{graphicx}
\usepackage[caption=false]{subfig}
\usepackage[colorlinks]{hyperref}
\usepackage[bottom]{footmisc}
\usepackage{enumerate}
\usepackage{float}
\usepackage{mathtools}

\usepackage{multirow}
\usepackage{colortbl}
\usepackage{lipsum}
\usepackage[most]{tcolorbox}

\usepackage{empheq}
\usepackage{tikz}
\usetikzlibrary{patterns,tikzmark, matrix,decorations.pathreplacing, calc, positioning,fit}

\usepackage{hf-tikz}

\usepackage[normalem]{ulem}
\DeclareMathAlphabet\mbc{OMS}{cmsy}{b}{n}

\usepackage{balance}

\usepackage{diagbox}

\usepackage{scalerel}

\begin{document}

\newcommand{\ks}[1]{{\textcolor{teal}{[KS: #1]}}}
\newcommand{\cj}[1]{{\textcolor{green}{[CJ: #1]}}}

\global\long\def\eqn#1{\begin{align}#1\end{align}}
\global\long\def\vec#1{\overrightarrow{#1}}
\global\long\def\ket#1{\left|#1\right\rangle }
\global\long\def\bra#1{\left\langle #1\right|}
\global\long\def\bkt#1{\left(#1\right)}
\global\long\def\sbkt#1{\left[#1\right]}
\global\long\def\cbkt#1{\left\{#1\right\}}
\global\long\def\abs#1{\left\vert#1\right\vert}
\global\long\def\cev#1{\overleftarrow{#1}}
\global\long\def\der#1#2{\frac{{d}#1}{{d}#2}}
\global\long\def\pard#1#2{\frac{{\partial}#1}{{\partial}#2}}
\global\long\def\re{\mathrm{Re}}
\global\long\def\im{\mathrm{Im}}
\global\long\def\dd{\mathrm{d}}
\global\long\def\ddd{\mathcal{D}}

\global\long\def\avg#1{\left\langle #1 \right\rangle}
\global\long\def\mr#1{\mathrm{#1}}
\global\long\def\mb#1{{\mathbf #1}}
\global\long\def\hmb#1{\hat{\mathbf #1}}
\global\long\def\mc#1{\mathcal{#1}}
\global\long\def\tr{\mathrm{Tr}}

\global\long\def\nth{$n^{\mathrm{th}}$\,}
\global\long\def\mth{$m^{\mathrm{th}}$\,}
\global\long\def\non{\nonumber}

\newcommand{\orange}[1]{{\color{orange} {#1}}}
\newcommand{\cyan}[1]{{\color{cyan} {#1}}}
\newcommand{\teal}[1]{{\color{teal} {#1}}}
\newcommand{\blue}[1]{{\color{blue} {#1}}}
\newcommand{\yellow}[1]{{\color{yellow} {#1}}}
\newcommand{\green}[1]{{\color{green} {#1}}}
\newcommand{\red}[1]{{\color{red} {#1}}}
\newcommand{\purple}[1]{{\color{purple} {#1}}}

\global\long\def\todo#1{\cyan{{$\bigstar$ \orange{\bf\sc #1 }}$\bigstar$} }

\global\long\def\addref#1{\orange{{$\bigstar$ \cyan{\bf\sc Add reference }}$\bigstar$} }

\global\long\def\redflag#1{\Rflag{first} \red{\bf \sc #1}}

\title{Fluctuation-induced Forces on Nanospheres in External Fields}

\author{Clemens Jakubec}
\email{clemens.jakubec@univie.ac.at}
\affiliation{Vienna Center for Quantum Science and Technology (VCQ), Faculty of Physics,
University of Vienna, Boltzmanngasse 5, A-1090 Vienna, Austria}

\author{Pablo Solano}
\affiliation{Departamento de F\'{i}sica, Facultad de Ciencias F\'{i}sicas y Matem\'{a}ticas, Universidad de Concepci\'{o}n, Concepci\'{o}n,
Chile}

\author{Uro\v{s} Deli\'c}
\affiliation{Vienna Center for Quantum Science and Technology (VCQ), Faculty of Physics,
University of Vienna, Boltzmanngasse 5, A-1090 Vienna, Austria}
\author{Kanu Sinha}
\email{kanu@arizona.edu}
\affiliation{Wyant College of Optical Sciences and Department of Physics, University of Arizona, Tucson, AZ USA 85721}

\begin{abstract}
We analyze the radiative forces between two dielectric nanospheres mediated via the quantum and thermal fluctuations of the electromagnetic field in the presence of an external drive. We generalize the scattering theory description of fluctuation forces to include external quantum fields,  allowing them to be in an arbitrary quantum state. The known trapping and optical binding potentials are recovered for an external coherent state. We demonstrate that an external squeezed vacuum state creates similar potentials to a laser, despite its zero average intensity. Moreover, Schr\"{o}dinger cat states of the field can enhance or suppress the optical potential depending on whether  they  are odd or even. Considering the nanospheres trapped by optical tweezers, we examine the total interparticle potential as a function of various experimentally relevant parameters, such as the field intensity, polarization, and phase of the trapping lasers. We demonstrate that an appropriate set of parameters could produce mutual bound states of the two nanospheres with potential depth as large as $\sim200$ K. Our results are pertinent to ongoing experiments with trapped nanospheres in the macroscopic quantum regime, paving the way for engineering interactions among macroscopic quantum systems.
\end{abstract}

\maketitle

\section{Introduction}
Bringing massive systems to the quantum regime is an essential step towards understanding the quantum-to-classical transition, a goal of foundational importance~\cite{Zurek91}.  Remarkable progress in the control of atomic, molecular, and optical (AMO) systems has made the macroscopic quantum regime increasingly accessible to experiments~\cite{Raimond2001,MQPRMP,  Arndt2014}: The Schr\"{o}dinger's cat  has been realized by photons, atoms and mechanical resonators~\cite{Deleglise2008, SCAtom, Vlastakis13, Bild2023}, molecular clusters as massive as $\sim 25000$ amu have been shown to exhibit quantum interference~\cite{Fein2019}, and millimeter-sized objects have been cooled down to their quantum ground states~\cite{Rossi2018}.

Optically levitated dielectric nanospheres are among the most promising experimental platforms for realizing large superpositions of massive objects~\cite{GonzalezBallestero21}. First pioneered by Ashkin  in 1970~\cite{Ashkin}, they bring together the advantages of optical trapping and cooling methods in terms of control, while being well-isolated from the environment without mechanical clamping, thus minimizing decoherence. Recent experiments have brought dielectric nanospheres as massive as $\sim 10^8 $ amu to their motional quantum ground states~\cite{Delic20, Marin2022, Piotrowski2023, Magrini2021, Tebbenjohanns2021, Kamba2022}. This finds application in quantum sensing and metrology while also providing an avenue for investigating foundational questions such as quantum-to-classical transition and gravitational interaction of quantum systems. More recent experiments have achieved tunable interactions between two silica nanospheres trapped via optical tweezers to create non-conservative and non-reciprocal interparticle potentials \cite{Rieser22}. The ability to precisely control and engineer the interactions between two dielectric nanospheres paves the way for realizing correlated macroscopic quantum systems.

 \begin{figure}[b]
    \centering
\includegraphics[width = 0.4\textwidth]{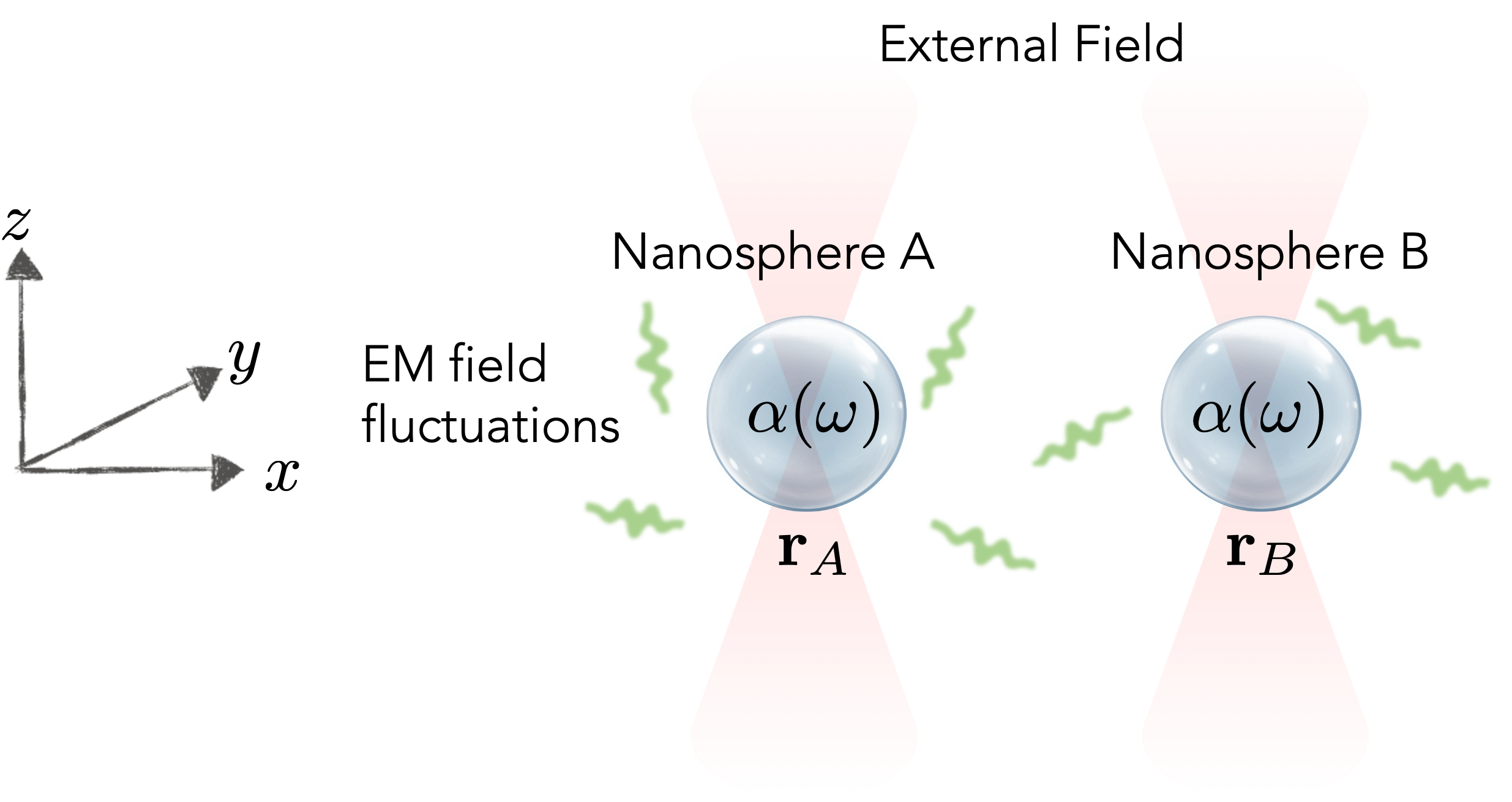}

    \caption{Schematic representation of two nanospheres (A and B) interacting via the quantum and thermal fluctuations of the EM field and externally applied arbitrary quantum fields. The spheres are located at $ \mb {r}_A $ and  $\mb{r}_B$, separated along the $x$-axis, with a polarizability $ \alpha\bkt{\omega}$ each.}
    \label{Fig:Sch}
\end{figure}

When considering interactions between two nanospheres in the near-field regime, one will inevitably encounter  fluctuation-induced forces resulting from the quantum and thermal fluctuations of the electromagnetic (EM) field \cite{Milonni}. Such fluctuation-induced  forces  exist even when the EM field is in the vacuum state  and are typically attractive in nature, thus imposing fundamental constraints on how close can two particles be stably trapped near each other and limiting the realization of macroscopic quantum states of such particles \cite{Rahi10} as well as influencing the decoherence of such states. It is thus critical to  develop experimentally amenable ways to control fluctuation-induced forces between nanoparticles in the near-field regime. Furthermore, the role of fluctuations in internal temperature sensing of levitated nanospheres has been investigated \cite{Agrenius2023}.

Previous works have shown that fluctuation-induced forces can be substantially modified in the presence of external drives \cite{MilonniSmith96, Chang14,  Fuchs18a, Fuchs18b, Sinha2020PRA}. Such drive-induced modifications to the vacuum forces can be significant compared to its pure fluctuation-induced counterpart, and even repulsive in character. Turning the strong short-ranged Casimir attraction to repulsion can allow for levitation and trapping particles at nanoscales   \cite{Chang14,Sinha18} and mitigate stiction in nano- and micro-mechanical devices \cite{Bostron12}. Therefore, control over external drives can  open new avenues for tailoring  fluctuation  forces  in a  system of two or more nanospheres.

In this work we explore fluctuation-induced forces in a system of two dielectric nanospheres in the presence of an externally applied field in a general quantum state. We describe the radiative interactions between the nanospheres mediated via the total field  by summing over the various  scattering processes to obtain the  potentials seen by each sphere for a general quantum state of the external field  to second-order in the particle polarizabilities. Analyzing these potentials for specific states of the external field, we show that in addition to the first and second order Casimir-Polder (CP) potential: (1) for a coherent state of the external field one recovers the well-known single particle trapping and interparticle optical binding potentials; (2)  a squeezed vacuum state of the external field can generate an equivalent potential to the trapping and optical binding potentials, even in the absence of a coherent amplitude; (3) a cat state of the external field can allow one to tune the trap and optical binding potentials via the phase between the two superposed coherent states.
We further analyze the scaling behavior of the various contributions to the total potential for nanospheres trapped in optical tweezers for different regimes of the interparticle separation. Our results show that combining the near-field fluctuation forces with optical binding can create mutual bound states of the two nanospheres with a potential depth as large as $\sim 200 $~K. We study such potentials as a function of various drive parameters -- intensity, polarization and the relative optical phase of the tweezer fields.

The paper is organized as follows. First, in Sec.~\ref{Sec:Model}, we present our model to describe the system up to second-order in the particle polariazbility $(\alpha(\omega))$, which we use to derive the interparticle potentials for general states of the external field in Sec.~\ref{Sec:GenStatePot}. We then study the obtained potentials in the presence of tweezer fields in Sec.~\ref{Engineering} and analyze their parameter dependence  in the light of finding bound states of two nanospheres, presenting a summary and outlook of our work in Sec.~\ref{Discussion}.

\section{Model}
\label{Sec:Model}

\begin{figure}
    
 \includegraphics[width = 0.4\textwidth]{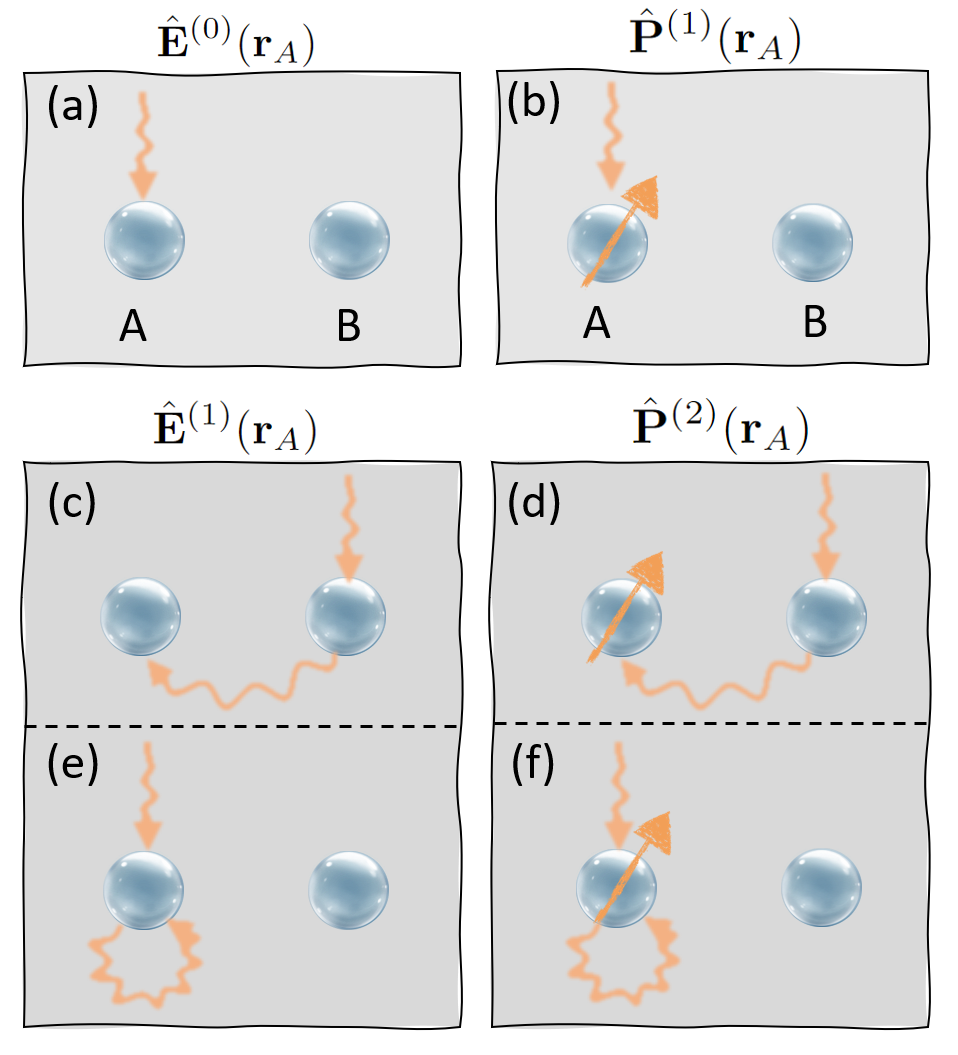}

    \caption{ Electric field and induced dipole moment at $\mb{r}_A$ to the lowest orders in the particle polarizabilities $\alpha(\omega)$ as denoted by  $\hat{\mb{E}}^{(0,1)}\bkt{\mb{r_A}}$ and $\hat{\mb{P}}^{(1,2)}\bkt{\mb{r_A}}$ at the position of sphere A. }
    \label{Fig:Sch2}
\end{figure}
We consider a system of two  nanospheres at positions $\mb{r}_A$ and $\mb{r}_B$, optically trapped by  external  fields, as shown in Fig.~\ref{Fig:Sch}. The Hamiltonian of the system is  given by $ \hat H = \hat H_\mr{F} +  \hat H_\mr{int} $, where:
\eqn{\hat{H}_F=&\sum_{\lambda=e,m}\int d^3\mb{r} \int_0^\infty d\omega \hbar\omega\hat{\mb{f}}_\lambda^\dagger(\mb{r},\omega)\cdot\hat{\mb{f}}_\lambda(\mb{r},\omega),
} is the Hamiltonian of the quantized EM field, with  $\hat{\mb{f}}_\lambda^{(\dagger)}\bkt{\mb{r}, \omega}$  as bosonic  operators for the EM field  in the macroscopic QED formalism~\cite{Buhmann1, Buhmann2}. These operators  obey the   canonical commutation relations $ \sbkt{\hmb{f}_\lambda\bkt{\mb{r}, \omega},\hmb{f}_{\lambda'}^\dagger\bkt{\mb{r}', \omega'} }=\delta (\omega - \omega')\delta \bkt{\mb{r} - \mb{r}'}\delta_{\lambda\lambda'} $. The dipole interaction Hamiltonian
\eqn{
\label{eq:int_H}
\hat {H}_\mr{int} =&- \sum_{i=A,B}\frac{1}{2} \hat{\mb{P}}(\mb{r}_i,t)\cdot\hat{\mb{E}}(\mb{r}_i,t)
}
represents the interaction between the induced dipole moments of the individual nanospheres $\hmb{P}\bkt{\mb{r}_i,t}$ and the electric field $\hmb{E}\bkt{\mb{r}_i,t}$ in the point-particle approximation. The induced dipole moment is defined as $\mb{P}=\alpha\mb{E}$. 

The total electric field seen by sphere $i$ is $\hmb{E}(\mb{r}_i,t)=\hat{\mb{E}}_\mr{f}(\mb{r}_i,t)+\hat{\mb{E}}_\mr{ex}(\mb{r}_i,t)$, where  $\hat{\mb{E}}_\mr{f}(\mb{r}_i,t)$ refers to  field fluctuations and $ \hmb{E}_\mr{ex}\bkt{\mb{r}_i,t}$ represents  an external electric field, such as that of a laser. Since the particles act as scatteres, light can bounce between the spheres multiple times. Each time the field will pick up an additional factor of the polarizability $\alpha$. Consequently, we can describe both the fluctuation field and the external field, including all their scattered components, as an expansion in $\alpha$. We express each of these fields up to first order in the particle polarizability $ \alpha (\omega)$, including the components scattered off the nanospheres:

\eqn{ \hat{\mb{E}}_\mr{f, {ex}}(\mb{r}_i,t)= \hat{\mb{E}}_\mr{{f}, {ex}}^{(0)}(\mb{r}_i,t)+ \hat{\mb{E}}_{\mr{f}, \mr{ex}}^{(1)}(\mb{r}_i,t)+O(\alpha^2),}

where the superscripts $k$ in   $\hmb{E}^{(k)}_{\mr{f}, \mr{ex}}$ refer to the order of polarizability $ \alpha^k (\omega)$. The zeroth order field

\eqn{\label{eq:E0}
\hat{\mb{E}}^{(0)}_\mr{f}(\mb{r}_i)=\sum_{\lambda=e,m}\int d^3\mb{r} \int_0^\infty d\omega G^{\lambda}(\mb{r}_i, \mb{r},\omega)\cdot\hat{\mb{f}}_\lambda(\mr{r},\omega)& \non\\
+\mr{H. c.}&
}
refers to the quantum and thermal fluctuations of the free electric field, as depicted in Fig.~\ref{Fig:Sch2}~(a).  The coefficients $  G_\lambda (\mb{r}, \mb{r}' ,\omega)$  are proportional to the Green's tensor  $  G\bkt{\mb{r}, \mb{r}',\omega}$ of the EM field, such that \cite{Buhmann1}:
\eqn{\label{eq:ggdag}
\sum_{\lambda }\int \dd^ 3 \mb{r}' {G}_\lambda \bkt{\mb{r}_1, \mb{r}', \omega}\cdot 
{G}^\dagger_\lambda \bkt{\mb{r}_2, \mb{r}', \omega} = \non\\
\frac{\hbar \mu_0 }{\pi }\omega^2 \im {G} \bkt{\mb{r}_1 ,\mb{r}_2, \omega},
}
where $ \lambda $ characterizes the source of the quantum noise polarization $(\lambda = e)$ or magnetization $(\lambda = m)$.  The Green's tensor $ {G} \bkt{\mb{r}_1 ,\mb{r}_2, \omega} $  describes the propagation of a photon at frequency $ \omega$ between positions $ \mb r_1 $ and $ \mb r_2$ and is obtained as a solution to the inhomogeneous  Helmholtz equation (see Appendix~\ref{app_A} for details)~\cite{Jackson}. The expression

\eqn{ \label{eq:E1}\hat{\mb{E}}_\mr{f}^{(1)}(\mb{r}_i)=&\sum_{j=A,B}\sum_{\lambda=e,m}\int d^3\mb{r} \int_0^\infty d\omega\alpha(\omega)\omega^2\mu_0\non\\
    & G(\mb{r}_i, \mb{r}_j,\omega)\cdot G^{\lambda}(\mb{r}_j, \mb{r},\omega)\cdot \hat{\mb{f}}_\lambda(\mr{r},\omega)+\mr{H.c.}
}
stands for the electric field at the position $ \mb{r}_i$ of sphere $i$, sourced by the dipole moment induced in sphere $j$ by the fluctuation field, as shown in Fig.~\ref{Fig:Sch2}~(c) and (e).  We assume a real polarizability $\alpha\bkt{\omega} $ of the nanospheres such that there is no internal dissipation. 

Similarly the incident and scattered external fields $\hat{\mb{E}}_\mr{ex}(\mb{r},t)$  seen by sphere $i$  are given by:

\begin{align}
\label{eq:E0ex}
    \hmb{E}_\mr{ex}^{(0)}(\mb{r}_i,t)=&\frac{1}{2}\sum_{\sigma}\int d^3\mb{k}\,\mb{\Phi}_\sigma(\mb{r}_i,\mb{k},\omega)\hat{a}_\sigma(\mb{k})e^{-i\omega t}+\mr{H.c.}\\    
    \label{eq:E1ex}
\hmb{E}_\mr{ex}^{(1)}(\mb{r}_i,t)=&\frac{1}{2}\sum_{\sigma}\int d^3\mb{k}\,\mu_0\omega^2\alpha(\omega)G(\mb{r}_i,\mb{r}_j,\omega)\cdot\non\\
&\quad\quad\quad\quad\mb{\Phi}_\sigma(\mb{r}_j,\mb{k},\omega)\hat{a}_\sigma(\mb{k})e^{-i\omega t}+\mr{H.c.}
\end{align}
Here $\mb{\Phi}_\sigma(\mb{r},\mb{k},\omega)=\sqrt{\hbar \omega/2\pi\epsilon_0(2\pi)^3}\Tilde{\mb{\Phi}}_\sigma(\mb{r},\mb{k},\omega)$ where $\Tilde{\mb{\Phi}}_\sigma(\mb{r},\mb{k},\omega)$ is the mode function of the field normalized to one at its maximum with wavevector $\mb{k}$, angular frequency $\omega$ and polarization $\sigma$.

The fields $\hmb{E}^{(k)}_\mr{f, ex} (\mb r_i , t)$ at sphere $i$ can each induce a  dipole moment in sphere $i$ given by $\hmb P_\mr{f , ex}^{(k+1)} (\mb r_i, t )$:
\eqn{
\label{eq:polarization}
 \hat{\mb{P}}_\mr{f,ex}(\mb{r}_i)= \hat{\mb{P}}_\mr{f,ex}^{(1)}(\mb{r}_i)+ \hat{\mb{P}}_\mr{f,ex}^{(2)}(\mb{r}_i)+ O(\alpha^3).
}
The first term in the above equation represents the dipole moment induced by the EM field fluctuations (f) or the incident external field (ex), as shown in Fig.~\ref{Fig:Sch2}~(b).  The second term refers to the dipole moment induced by the electric field scattered off one of the spheres, as in Fig.~\ref{Fig:Sch2}~(d) and (d).

The fluctuating dipole moment up to second order in $\alpha (\omega)$ is
\eqn{
\label{eq:p1}
&\hat{\mb{P}}_\mr{f}^{(1)}(\mb{r}_i)=\non\\
&\sum_{\lambda=e,m}\int d^3\mb{r} \int_0^\infty d\omega\alpha(\omega)G^{\lambda}(\mb{r}_i, \mb{r},\omega)\cdot\hat{\mb{f}}_\lambda(\mb{r},\omega) +\mr{H. c.}\\
\label{eq:p2}
&\hat{\mb{P}}_\mr{f}^{(2)}(\mb{r}_i)=\sum_{j=A,B}\sum_{\lambda=e,m}\int d^3\mb{r} \int_0^\infty d\omega\alpha(\omega)^2\omega^2\mu_0& \non\\
&\quad\quad\quad\quad G(\mb{r}_i, \mb{r}_j,\omega)\cdot G^{\lambda}(\mb{r}_j, \mb{r},\omega)\cdot \hat{\mb{f}}_\lambda(\mb{r},\omega)+\mr{H. c.},
}

The dipole moment induced in sphere $i$ by the external field 
 $ \hmb{E}_\mr{ex}(\mb r _i , t)$ can be similarly obtained as:
\eqn{\label{eq:p1ex}
    \mb{P}_\mr{ex}^{(1)}(\mb{r}_i,t)=\frac{1}{2}\sum_{\sigma}\int d^3\mb{k}\,\alpha(\omega)\mb{\Phi}_\sigma(\mb{r}_i,\mb{k},\omega)\hat{a}_\sigma(\mb{k})e^{-i\omega t}&\non\\
    +\mr{H.c.}&
    }
    \eqn{\label{eq:p2ex}
    \mb{P}_\mr{ex}^{(2)}(\mb{r}_i,t)=&\frac{1}{2}\sum_{\sigma}\int d^3\mb{k}\,\alpha(\omega)^2\mu_0\omega^2G(\mb{r}_i,\mb{r}_j,\omega)\cdot\non\\
&\quad\quad\quad\quad\mb{\Phi}_\sigma(\mb{r}_j,\mb{k},\omega)\hat{a}_\sigma(\mb{k})e^{-i\omega t}+\mr{H.c.}
}

We can thus combine the contributions to second order in polarizability $\alpha$ and write the interaction Hamiltonian in Eq.~\eqref{eq:H_int} as:

\begin{align}
\begin{split}
\label{eq:H_int}
    \hat{H}_\mr{int}=\hat{H}_\mr{int}^{(1)}&+\hat{H}_\mr{int}^{(2)}+O(\alpha^3),
\end{split}
\end{align}
where the first-order interaction Hamiltonian
\eqn{
\label{eq:H_int^(1)}
\hat{H}_\mr{int} ^{(1) }\equiv &\sum_{i=A,B}-\frac{1}{2}\hat{\mb{P}}^{(1)}(\mb{r}_i,t)\cdot\hat{\mb{E}}^{(0)}(\mb{r}_i,t),
}
represents the interaction between the electric field at the position of the nanospheres and the dipole moment it induces, as shown by  Fig.~\ref{Fig:Sch2}~(a) and (b). We note that the contributing processes only pertain to a single sphere, thus $ H^{(1) }_\mr{int}$ does not contribute to the interparticle potential.

The second-order Hamiltonian is given by:

\eqn{
\label{eq:H_int^(2)}
 \hat{H}_\mr{int} ^{(2) }\equiv \sum_{i=A,B}-\frac{1}{2}\hat{\mb{P}}^{(1)}(\mb{r}_i,t)\cdot\hat{\mb{E}}^{(1)}(\mb{r}_i,t)&\non\\
 -\frac{1}{2}\hat{\mb{P}}^{(2)}(\mb{r}_i,t)\cdot\hat{\mb{E}}^{(0)}(\mb{r}_i,t),&
}
where the first term corresponds to the interaction between the dipole moment induced by the fluctuation or external  field in sphere $i$  (Fig.~\ref{Fig:Sch2}~(b)) and the electric field scattered off either of the particles at position $ \mb{r}_i$ (Fig.~\ref{Fig:Sch2}~(c) and (e)). The second  term corresponds  to the interaction between the dipole moment induced in particle $i$ by the field scattered off of one of the  particles (Fig.~\ref{Fig:Sch2}~(d) and (f)) and the fluctuation or external fields at $ \mb{r}_i $ (as in Fig.~\ref{Fig:Sch2}~(a)).

\section{Potentials for general states of the field}
\label{Sec:GenStatePot}

 We now focus on the potential that arises from the fields and interactions described in the previous section. We denote the total potential seen by sphere $ i $ as \eqn{U_i (\mb r_i, \mb r_j) =U_{i}^{(1)} (\mb r_i)   +U_{i}^{(2)} (\mb r_i, \mb r_j )    ,} where $U_{i}^{(1)} (\mb r_i, \mb r_j)  = \tr_\mr{F} \sbkt{\rho_\mr{F} \hat H_\mr{int}^{(1)}} $  
 and $U_{i}^{(2)} (\mb r_i, \mb r_j)  = \tr_\mr{F} \sbkt{\rho_\mr{F} \hat H_\mr{int}^{(2)}} $  represent the first and second-order potentials  in the particle polarizibility calculated in first-order perturbation theory. It suffices to include first-order shifts because the Hamiltonians Eq.~\eqref{eq:H_int^(1)} and \eqref{eq:H_int^(2)}  include all scattering processes to second-order in the particle polarizability.   We analyze these potentials below for various quantum states of total field denoted by $\rho_\mr{F}=\rho_\mr{th}\otimes \rho_\mr{ex}$, with $\rho_\mr{th}=\mr{exp}(-\hat{H}_\mr{F}/k_BT)/Z$  being the thermal state of the fluctuation field, with  $Z$ as the partition function, and $\rho_\mr{ex}$ being a general  state of the external field.  

\subsection{First-order potential}
\label{First-order potential}

We find the first-order potential seen by the sphere A as $U ^{(1) }_A (\mb r_A) = U_{A,\mr{f}}^{(1) }(\mb{r}_A)+ U_{A,\mr{ex}}^{(1) }(\mb{r}_A)$,  where \eqn{\label{eq:U1f}U_{A,\mr{f}}^{(1) }(\mb{r}_A) = -\frac{1}{2}\tr_\mr{F}\sbkt{\rho_\mr{F} \hmb{P}_\mr{f} ^{(1) } (\mb{r}_A, t )\cdot \hmb{E}_\mr{f} ^{(0) } (\mb{r}_A, t ) }} represents the contribution from quantum and thermal fluctuations of the field  and \eqn{\label{eq:U1ex}U_{A,\mr{ex}}^{(1) }(\mb{r}_A) =-\frac{1}{2}\tr_\mr{F}\sbkt{\rho_\mr{F} :\hmb{P}_\mr{ex} ^{(1) } (\mb{r}_A, t ) \cdot\hmb{E}_\mr{ex} ^{(0) } (\mb{r}_A, t ): }} is the potential  induced by the  external field. We note that the cross-coupling terms between the fluctuation field and the external field that are linear in  $ \cbkt{\hmb{f}_\lambda, \hmb{f}^\dagger_\lambda} $, vanish in first order perturbation theory, as $\tr\sbkt{\hat{\rho}_\mr{F}\hat{\mb{f}}^{(\dagger)}_\lambda(\mb{r},\omega)}=0$ for diagonal states like the thermal state, but can be non-zero for other states. 
 Furthermore, the Hamiltonian $ H _\mr{int}^{(1) }$ is normal ordered with respect to the operators $\hat{a}_\sigma(\mb{k})$ and $\hat{a}^\dagger_\sigma(\mb{k})$ of the external field. In our description fluctuation effects arise solely from $\hat{\mb{E}}_\mr{f}(\mb{r}_i,t)$. Nevertheless, $\hat{\mb{E}}_\mr{ex}(\mb{r}_i,t)$ is a quantum field, which exhibits ground state fluctuations. Consequently, the normal ordering is necessary to prevent the overcounting of ground state effects.   
 Substituting  the  field $ \hmb{E}_{\mr{f}}^{(0)}$ (Eq.~\eqref{eq:E0}) and  dipole moment $ \hmb{P}_{\mr{f}}^{(1)}$ (Eq.~\eqref{eq:p1}) in Eq.~\eqref{eq:U1f}  we obtain the first-order fluctuation-induced potential:  

\eqn{\label{eq:UAF1}
&U_{A,\mr{f}}^{(1)}(\mb r_A) =\non\\
&-\frac{\hbar \mu_0 }{2\pi }\int d\omega~ \alpha(\omega)\omega^2(2n(\omega)+1)\tr \sbkt{\im \ G_{AA}(\omega)},}
where $n(\omega)=\frac{1}{e^{{\hbar\omega}/\bkt{k_BT}}-1}$ is the average thermal photon number and we have defined the shorthand notation $G_{ii}(\omega)\equiv G(\mb{r}_i,\mb{r}_i,\omega)$.  If there are no surfaces present in the system, then the Green's tensor is given by the free-space Greens tensor. The imaginary part of the free-space Green's tensor, which in the coincidence limit is given by $\im \ G_\mr{free}(\mb{r}_i,\mb{r}_i,\omega)=\mathbb{1}k/6\pi$, yields a constant potential given by:

\begin{align}
    U_{A,\mr{f}}^{(1)}(\mb r_A) =-\frac{\hbar\mu_0}{4\pi^2c}\int d\omega~ \alpha(\omega)\omega^3(2n(\omega)+1)
\end{align}

However, if there is a boundary present in the system, the Green's tensor can be split into a free part and a scattering part, $G(\mb{r}_i,\mb{r}_j,\omega)=G_\mr{free}(\mb{r}_i,\mb{r}_j,\omega)+G_\mr{sc}(\mb{r}_i,\mb{r}_j,\omega)$. In the coincidence limit, the scattering part depends on the distance between the nanosphere and the surface. Consequently, the single particle potential will only give a distance-dependent energy, or a force, if there is a boundary  present in the system, corresponding to the usual thermal Casimir-Polder potential~\cite{Buhmann2}.

Similarly, we obtain the first-order potential induced by the external field by substituting Eq.~\eqref{eq:E0ex} and \eqref{eq:p1ex} in \eqref{eq:U1ex}: 
\eqn{
\label{eq:UAEX1}
&U_{A,\mr{ex}}^{(1)}(\mb r_A) =-\int d^3\mb{k}\int d^3\mb{k}'\alpha(\omega)\non\\
&\re \ \cbkt{\sum_{\sigma\sigma'}\mb{\Phi}_{\sigma'}^{A\dagger}\cdot \mb{\Phi}_{\sigma}^A\avg{ \hat{a}^\dagger_\sigma(\mb{k})\hat{a}_{\sigma'}(\mb{k'})} e^{-i(\omega-\omega')t}},
}
where we have used the abbreviations $\mb{\Phi}_{\sigma}^{i}\equiv \mb{\Phi}_{\sigma}(\mb{r}_i,\mb{k},\omega)$, $\mb{\Phi}_{\sigma'}^{i}\equiv \mb{\Phi}_{\sigma'}(\mb{r}_i,\mb{k'},\omega')$ . If the external field is a single-mode field, the above potential is proportional to the expectation number of photons in that mode.

\subsection{Second-order potential}

The second-order potential seen by sphere A can be obtained from Eq.~\eqref{eq:H_int^(2)} as $ U_A^{(2) }(\mb{r}_A,\mb{r}_B) = U_{A,\mr{f}}^{(2) }(\mb{r}_A,\mb{r}_B) + U_{A,\mr{ex}}^{(2) }(\mb{r}_A,\mb{r}_B) $, where:
 
\eqn{\label{eq:UA2f}
   & U_{A,\mr{f}}^{(2) }(\mb{r}_A,\mb{r}_B)=-\frac{1}{2}\tr_\mr{F}\sbkt{\rho_\mr{F} \cbkt{\hmb{P }^{(1) }_\mr{f}(\mb r_i , t)\cdot \hmb{E}^{(1)}_\mr{f}(\mb r_i , t) \right.\right.\non\\   &\quad\quad\quad\quad\quad\quad\quad\quad\quad\left.\left.+\hmb{P }^{(2) }_\mr{f}(\mb r_i , t)\cdot \hmb{E}^{(0)}_\mr{f}(\mb r_i , t)  }}}
   represents the fluctuation-induced component, and 
 \eqn{ \label{eq:UA2ex}& U_{A,\mr{ex}}^{(2) }(\mb{r}_A,\mb{r}_B)=-\frac{1}{2}\tr_\mr{F}\sbkt{\rho_\mr{F} \cbkt{:\hmb{P }^{(1) }_\mr{ex}(\mb r_i , t)\cdot \hmb{E}^{(1)}_\mr{ex}(\mb r_i , t): \right.\right.\non\\
&\quad\quad\quad\quad\quad\quad\quad\quad\left.\left.+:\hmb{P }^{(2) }_\mr{ex}(\mb r_i , t)\cdot \hmb{E}^{(0)}_\mr{ex}(\mb r_i , t)  }:},}
the component induced by the external field. Substituting the fields (Eq.~\eqref{eq:E0} and \eqref{eq:E1}) and induced dipole moments (Eq.~\eqref{eq:p1} and \eqref{eq:p2}) in Eq.~\eqref{eq:UA2f}, yields the fluctuation-induced second-order potential for sphere A:

\eqn{\label{eq:general_second_order}
    U_{A,\mr{f}}^{(2)}(\mb r_A ,\mb r_B)= &-\frac{\hbar\mu_0^2}{\pi}\int d\omega ~\alpha(\omega)^2\omega^4(2n(\omega)+1)\non\\
    &\sbkt{\tr \sbkt{\im \ G_{AA}(\omega)\cdot \re \ G_{AA}(\omega)}\right.\non\\
    & \left.+\tr \sbkt{ \im \ G_{AB}(\omega)\cdot \re \ G_{BA}(\omega)}},
    }
    where the first term represents the modification to the single-sphere potential arising from the self-interaction terms and can be ignored in the absence of external boundary conditions. The second term represents the interparticle thermal CP potential to second order in the particle polarizability~\cite{Emig07}. As before, if the substitute the free-space Green's tensor, thus neglecting the self-interactions, we get:
    \eqn{
    &U_{A,\mr{f}}^{(2)}(\mb r_A ,\mb r_B)= -\frac{c^4\hbar\mu_0^2}{16\pi^3r^6}\int d\omega ~(2n(\omega)+1)\sbkt{\text{sin}(2kr)\right.\non\\
&\left.\bkt{-3+5(kr)^2-(kr)^4}+\text{cos}(2kr)\bkt{-6+2(kr)^3}},
    }
    where $r=\abs{\mb{r}_A-\mb{r}_B}$. Similarly the externally-induced second order potential can be obtained from substituting the external fields (Eqs.~\eqref{eq:E0ex} and \eqref{eq:E1ex}) and  the corresponding induced dipole moments (Eqs.~\eqref{eq:p1ex} and \eqref{eq:p2ex}) in Eq.~\eqref{eq:UA2ex}:

\begin{widetext}
    \eqn{
    &U_{A,\mr{ex}}^{(2)}(\mb r_A ,\mb r_B)=\non\\
    &-\int d^3\mb{k}\int d^3\mb{k}'\,\mu_0\omega^2\alpha(\omega)\alpha(\omega')\re \ \cbkt{\sum_{\sigma\sigma'}\bkt{\mb{\Phi}_{\sigma'}^{A\dagger}\cdot G^\dagger_{AA}(\omega')\cdot \mb{\Phi}^A_{\sigma}+\mb{\Phi}_{\sigma'}^{A\dagger}\cdot G_{AA}(\omega)\cdot \mb{\Phi}^A_{\sigma}}\avg{ \hat{a}^\dagger_\sigma(\mb{k})\hat{a}_{\sigma'}(\mb{k'})} e^{-i(\omega-\omega')t}}\non\\
     &-\int d^3\mb{k}\int d^3\mb{k}'\,\mu_0\omega^2\alpha(\omega)\alpha(\omega')\re \ \cbkt{\sum_{\sigma\sigma'}\bkt{\mb{\Phi}_{\sigma'}^{A\dagger}\cdot G^\dagger_{AB}(\omega')\cdot \mb{\Phi}^A_{\sigma}+\mb{\Phi}_{\sigma'}^{A\dagger}\cdot G_{AB}(\omega)\cdot \mb{\Phi}^A_{\sigma}}\avg{ \hat{a}^\dagger_\sigma(\mb{k})\hat{a}_{\sigma'}(\mb{k'})} e^{-i(\omega-\omega')t}}
}
\end{widetext}
The first  and second terms in Eq. \ref{eq:p2ex} correspond to the modification of the single-sphere and the interparticle potentials resulting from the external field.  Both the single-particle and interparticle potential between two dielectric spheres for a general state of the external EM field are determined  by the expectation value $ \avg{\hat a _\sigma(\mb k)^\dagger\hat a _{\sigma'}(\mb k')}$. In the following we analyze particular cases of external field states.

\subsection{Coherent state}
\label{Sec:Coh}

We first consider the external field to be in a single-mode coherent state, corresponding to the case of optical tweezers. To this end, we apply the displacement operator $\hat{D}=e^{\beta_\sigma(\mb{k})\hat{a}_\sigma^\dagger(\mb{k})-\beta_\sigma^*\hat{a}_\sigma(\mb{k})}$ to the vacuum state of the total field,  giving an expectation value $\avg{ \hat{a}_\sigma(\mb{k})^\dagger\hat{a}_{\sigma'}(\mb{k'})}=\beta_\sigma(\mb k)^*\beta_\sigma(\mb k')$ .

 Since the mode function describes the spatial part of the electric field, we can map it to the electric field, thus showing that the coherent state potential reduces to the optical trap  and optical binding potential. The details of this calculation are presented in Appendix~\ref{App_B}. As a consequence, we find that the first-order external potential in this case  reduces to the well-known trap potential created by the optical tweezers after the application of a single-mode approximation at frequency $\omega_0$:

\begin{align}
\label{eq:single_mode_tweezer}
    U_{A,\mr{coh}}^ {(1)}(\mb{r}_A)=-\frac{1}{4}\alpha(\omega_0)\abs{\mb E_{\mr{tw}}(\mb r_A)}^2
\end{align}

Similarly, it can be seen that the second-order potential created by the tweezers  corresponds to the interparticle optical binding potential  \cite{Pavel2010, Burns1989, rudolph2023} (see Appendix~\ref{App_B}):

\begin{align}
\label{eq:single_mode_OB}
    &U_{A,\mr{ex}}^{(2)}(\mb r_A ,\mb r_B)=\non\\
    &-\frac{1}{2}\mu_0\omega_0^2\alpha(\omega_0)^2\re \ \cbkt{\mb E^*_{\mr{tw}}(\mb r_A)\cdot G(\mb r_A,\mb r_A,\omega_0)\cdot \mb E_{\mr{tw}}(\mb r_A)}\non\\
    &-\frac{1}{2}\mu_0\omega_0^2\alpha(\omega_0)^2\re \ \cbkt{\mb E^*_{\mr{tw}}(\mb r_A)\cdot G(\mb r_A,\mb r_B,\omega_0)\cdot \mb E_{\mr{tw}}(\mb r_B)},
\end{align}

Using the free-space Green's tensor as before and ignoring the divergent self-interaction term, the optical binding potential yields:  

\eqn{
\label{eq:free_space_OB}
U_{A,\mr{ex}}^{(2)}(\mb r_A ,\mb r_B)= \frac{c^2\mu_0\alpha(\omega_0)^2}{8\pi r^3}\re  \sbkt{h(kr)\mb E^*_{\mr{tw}}(\mb r_A)\cdot\mb E_{\mr{tw}}(\mb r_B)\right.\non\\
\left.-f(kr)\sum_{ij}\mb E^*_{\mr{tw}}(\mb r_A)_i \frac{r_i r_j}{r^2} \mb E_{\mr{tw}}(\mb r_B)_j}
}
where $h(kr)=e^{ikr}\sbkt{1-ikr-(kr)^2}$ and $f(kr)=e^{ikr}\sbkt{3-3ikr-(kr)^2}$. 
\subsection{Squeezed vacuum state}
\label{Sec:Sq}

We now consider the case where the external field is in a squeezed vacuum state for a single mode. The effect of squeezed light in a single particle system has been investigated in \cite{Romero-Isart_Squeezed}. The external field state is obtained by applying  the  single-mode squeezing operator $\hat{S}=e^{\frac{1}{2}(\xi_\sigma(\mb{k})\hat{a}_\sigma(\mb{k})^2-\xi_\sigma(\mb{k})\hat{a}_\sigma(\mb{k})^{\dagger 2})}$ to the ground state, with $\xi_\sigma(\mb{k})=r_\sigma(\mb{k})e^{i\Phi_\sigma(\mb{k})}$ as the squeezing parameter, where $\xi_\sigma(\mb{k})$ is defined such that it is non-zero only for a single mode $\mb{k}_0$. Together with the expectation value $\avg{ \hat{a}_\sigma(\mb{k})^\dagger\hat{a}_{\sigma'}(\mb{k'})}=\delta(\mb{k}-\mb{k}_0)\delta(\mb{k}'-\mb{k}_0)\delta_{\sigma\sigma'}{\text{sinh}^2r_\sigma(\mb{k})}$, this leads to the first and second-order potentials for sphere A as:

\eqn{
\label{eq:squeezed_1st_order}
&   U_{A,\mr{sq}}^ {(1)}(\mb{r}_A)=- \alpha(\omega) \mb{\Phi}_{\sigma}^{A\dagger}\cdot \mb{\Phi}_{\sigma}^A
\text{sinh}^2r_\sigma(\mb{k}_0)\\
&   U^{(2)}_{A,\mr{sq}}(\mb{r}_A,\mb{r}_B)=- \mu_0\omega^2\alpha(\omega)^22\text{sinh}^2r_\sigma(\mb{k}_0)\non\\
    &\bkt{\re \ \cbkt{\mb{\Phi}_{\sigma}^{A\dagger}\cdot G_{AA}(\omega)\cdot \mb{\Phi}^A_{\sigma}}
     +\re \ \cbkt{\mb{\Phi}_{\sigma}^{A\dagger}\cdot G_{AB}(\omega)\cdot \mb{\Phi}^B_{\sigma}}}
}
As a striking result, we observe in this case that the squeezed vacuum state creates: (1) a single particle potential similar to that created by a tweezer field (Eq.~\eqref{eq:U1coh}) and (2) an interparticle potential similar to the optical binding potential (Eq.~\eqref{eq:U2coh}).  To quantitatively compare the two, e.g., to see how much squeezing of the vacuum is necessary to produce an interaction of the same strength as the tweezer or the coherent optical binding potentials, we consider  the ratio of the  potentials $\frac{ U_{A,\mr{ sq}}^{(1) }}{U_{A, \mr{coh}}^{(1)}} =\frac{ U_{A,\mr{ sq}}^{(2) }}{U_{A, \mr{coh}}^{(2)}}$. As mentioned above, we assume the tweezer field to be single-mode and have a flat intensity profile such that only the $\omega_0$ mode of the vacuum is squeezed and only the $\omega_0$ mode is displaced, with no spatial dependence. Because the spatially dependent part under these approximations is the same for both potentials, we define the following ratio of the two potentials: $\frac{U_\mr{Sq}}{U_\mr{OB}}=\frac{2\text{sinh}^2r\abs{\Phi_\mr{sq}}^2}{\abs{\beta(\mb{k)}}^2\abs{\Phi_\mr{coh}}^2}=\frac{\text{sinh}^2r\hbar\omega_0c}{(2\pi)^4I}$, where $I$ is the intensity of the tweezer and we used that $\Tilde{\Phi}_\mr{sq}=1$. Using $I=10^{-2} W/\mu m^2$ and $\lambda_0=1064$ nm, the squeezing parameter would need to be $r\approx 27$, which amounts to a squeezing of $\approx 240$ dB for the squeezed state potential to be comparable to the coherent state potential.

\subsection{Cat state}
\label{Sec:Cat}

Another example we present is the case where the external field is in a Schr\"{o}dinger cat state. Cat states are defined as superpositions of coherent states given by $\ket{C}=N\bkt{\ket{\beta}+e^{i\theta}\ket{-\beta}}$, where $\ket{\beta}$ is a coherent state with coherent amplitude $\beta$ and $N=\sbkt{2+2\mr{exp}(-2\beta^2)\mr{cos}\theta}^{\frac{1}{2}}$ is the normalization factor. We consider a single-mode cat state, meaning that one mode $\omega_0$ is in a cat state and all other modes remain in the ground state. Using the single-mode cat state $\ket{C}$ as defined above, we find $\bra{C}\hat{a}_\sigma(\mb{k})^\dagger\hat{a}_{\sigma'}(\mb{k'})\ket{C}=N(\mb{k})N(\mb{k}')\beta^*(\mb{k})\beta(\mb{k}')\bkt{2-e^{i\theta(\mb{k})}-e^{-i\theta(\mb{k}')}}\delta(\mb{k}-\mb{k}_0)\delta(\mb{k}'-\mb{k}_0)$. Substituting these expectation values into Eqs. (\ref{eq:UAEX1}) and (\ref{eq:UA2ex}) and evaluating both integrals yields:

\eqn{
&U_{A,\mr{ex}}^{(1)}(\mb r_A) =\non\\
&-\alpha(\omega_0)\re \ \cbkt{\mb{\Phi}^{A\dagger}\cdot \mb{\Phi}^AN^2(\mb{k}_0)\abs{\beta(\mb{k}_0)}^22\bkt{1-\mr{cos}\theta}}
}
and
\eqn{
&U_{A,\mr{ex}}^{(2)}(\mb r_A ,\mb r_B)=-4\mu_0\omega_0^2\alpha(\omega_0)^2\non\\
&\re \ \biggl\{\mb{\Phi}^{A\dagger}\cdot\bkt{G_{AA}^\dagger(\omega_0)+G_{AB}^\dagger(\omega_0)}\cdot\mb{\Phi}^AN^2(\mb{k}_0)\non\biggr.\\
&\biggl.\abs{\beta(\mb{k}_0)}^2(1-\mr{cos}\theta)\biggr\}
}

If we compare this with the coherent state potentials for a single mode, we see that the cat state potential differs by a factor of $N^2(\mb{k}_0)(1-\mr{cos}\theta)$. Since $1-\mr{cos}\theta$ takes values between $0$ and $2$, it is possible to turn off the interaction potential as well as the trap potential by choosing a suitable $\theta$. This means that for an even cat state ($\theta=2n\pi$) there will be no potential in spite of the presence of an external laser field.  While for an odd cat state ($\theta=(2n+1)\pi $) the potential is enhanced by a factor of two compared to a coherent state with the same average photon number.

Furthermore, the potential created by a general statistical mixture $\hat{\rho}=p_1\ket{\beta}\bra{\beta}+p_2\ket{-\beta}\bra{-\beta}$ of coherent states with $p_1+p_2=1$ is equivalent to the potential created by a coherent state $\ket{\beta}$. This means that the potential of a single-mode cat state differs from that of a simple statistical mixture. The fact that both the first order and second order potentials are enhanced or suppressed for cat states can thus be solely attributed to the quantum coherence between the superposed states. 

\section{Engineering sphere-sphere potential}
\label{Engineering}

In this section we will focus on  engineering  the interparticle potential for the case where  two silica nanospheres are trapped via optical tweezers. We examine the optical potential by tuning various parameters: polarization, relative optical phase and intensity of the tweezers.

\subsection{Scaling regimes}
\label{sec:scaling regimes}

\begin{figure}[t]
    \centering
    \includegraphics[width=0.48\textwidth]{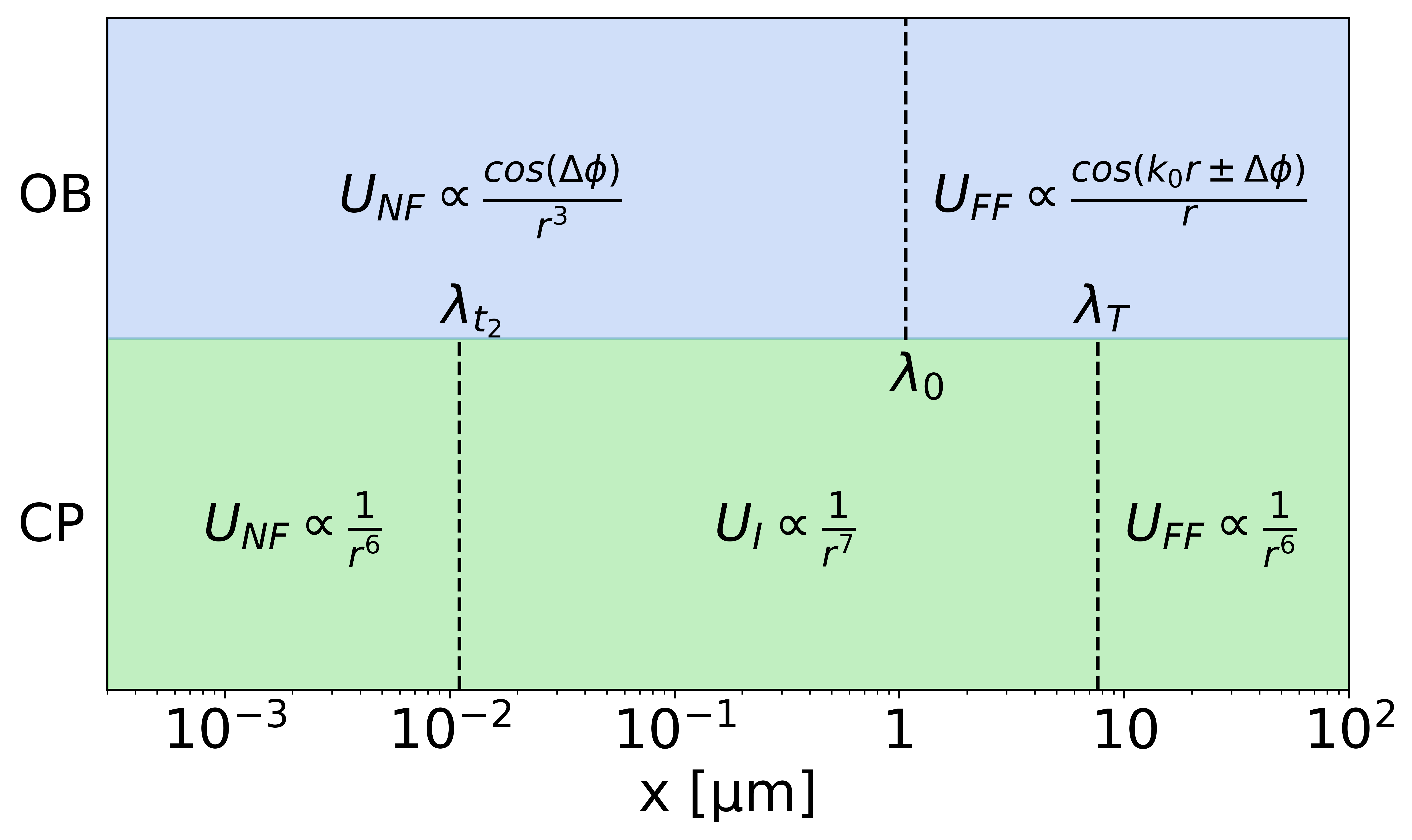}
    \caption{Comparison of all regimes for the optical binding and thermal Casimir-Polder potentials of two nanospheres. The optical binding potential (OB) scales as $\propto\mr{cos}(k_0r\pm \Delta\phi)/r^3$ in the near-field $(r\ll\lambda_0)$ and as $\propto\mr{cos}(k_0r\pm\Delta\phi)/r$ in the far-field $(r\gg\lambda_0)$. The Casimir-Polder potential (CP) scales as $\propto1/r^6$ in the near-field $(r<\lambda_{t_2})$, as $\propto1/r^7$ in the intermediate regime $(\lambda_{t_2}<r<\lambda_T)$ and as $\propto 1/r^6$ in the far-field $(r>\lambda_T)$.The length scales $\lambda_0,\lambda_{t_2}$ and $\lambda_T$ are defined in Sec. \ref{sec:scaling regimes}.}
    \label{fig:scaling_regimes}
\end{figure}

Before we consider the total potential in more detail, it is instructive to look at the scaling of the thermal Casimir-Polder and the optical binding potentials across various regimes of the sphere-sphere separation. A summary of the different scaling regimes is presented in Fig. \ref{fig:scaling_regimes}. 

\subsubsection{Thermal Casimir-Polder potential}
\label{sec:thermal_CP_approx}
The thermal Casimir-Polder potential exhibits three different scaling regimes, given by the characteristic length scales $\lambda_{t_2}$ and $\lambda_T$, where $\lambda_{t_2}= 11.3\ \mr{nm}$ is the dominant transition wavelength of silica, which appears in the Drude-Lorentz permittivity \cite{Boyer1975,Milonni1996, Spagnolo2007}. 

\begin{align}
    \epsilon(\omega)=1+\frac{\omega^2_{p_1}}{\omega_{t_1}^2-\omega^2-i\gamma_1\omega}+\frac{\omega^2_{p_2}}{\omega_{t_2}^2-\omega^2-i\gamma_2\omega}
\end{align}
in the form of a transition frequency $\omega_{t_2}$. The other parameters are the plasma frequencies $\omega_{p_1}=1.75\times10^{14}$ Hz, $\omega_{p_2}=2.96\times10^{16}$Hz, the transition frequencies $\omega_{t_1}=1.32\times10^{14}$ Hz, $\omega_{t_2}=2.72\times10^{16}$ Hz and the damping coefficients $\gamma_{1}=4.28\times10^{13}$ Hz, $\gamma_{2}=8.09\times10^{15}$ Hz \cite{PhysRevA.94.023621}. $\lambda_T=\frac{\hbar c}{k_B T}$ is the thermal length scale. The permittivity enters the polarizability $\alpha(\omega)$ via the Clausius-Mossotti relation $\alpha(\omega)=4\pi\epsilon_0 R^3\frac{\epsilon(\omega)-1}{\epsilon(\omega)+2}$,  $R$ being the radius of the nanospheres.

Introducing the dimensionless polarizability $\Tilde{\alpha}(\omega)=\frac{\epsilon(\omega)-1}{\epsilon(\omega)+2}$, we can write the thermal Casimir-Polder potential approximated in the near-field ($r\ll\lambda_{t_2}$) as \cite{Boyer1975}:

\begin{align}
\label{eq:thermal_CP_near}
     U_\mr{CP}^{\text{ts}}(\mb{r}_A, \mb{r}_B)\approx -4k_BT\bkt{\frac{R}{r}}^6\Tilde{N}_T-\frac{\hbar\omega_{t_2}}{\pi}\bkt{\frac{R}{r}}^6\Tilde{N}_0
\end{align}
where the dimensionless quantities $\Tilde{N}_T$ and $\Tilde{N}_0$ are $\Tilde{N}_T=\sum_j \Tilde{\alpha}(i\Tilde{\xi}_j)^2$ and $\Tilde{N}_0=\int_0^{\infty}d\Tilde{\omega}\Tilde{\alpha}(\Tilde{\omega})^2$,  with the sum being evaluated over the dimensionless Matsubara frequencies $\Tilde{\xi}_j = 2 \pi j \frac{k_B T  }{\hbar\omega_{t_2}}$. The first term corresponds to the thermal contribution with the thermal energy scale $k_BT$, whereas the second terms corresponds to the ground state contribution with energy scale $\hbar\omega_{t_2}$.

In the intermediate regime ($\lambda_{t_2}<r<\lambda_T$) the thermal Casimir-Polder potential can be approximated as \cite{Boyer1975}:

\begin{align}
\label{eq:thermal_CP_intermediate}
    U_\mr{CP}^{\text{ts}}(\mb{r}_A, \mb{r}_B)\approx-\frac{23}{4}\frac{\hbar c}{r \pi}\bkt{\frac{R}{r}}^6 \Tilde{\alpha}(0)^2,
\end{align}

with the characteristic energy scale as $\frac{\hbar c}{r}$ and in the far-field regime ($r\gg\lambda_T$) as:

\begin{align}
\label{eq:thermal_CP_far}
    U_\mr{CP}^{\text{ts}}(\mb{r}_A, \mb{r}_B)\approx-6k_BT\bkt{\frac{R}{r}}^6\Tilde{\alpha}(0)^2
\end{align}
with the thermal energy scale $k_BT$.

In the limit of $T\rightarrow0$, the thermal length scale $\lambda_T$ goes to infinity. The intermediate regimes will thus extend to infinity, revealing the two scaling regimes of the ground state Casimir-Polder potential between two dielectric spheres \cite{Boyer1975}. 
\begin{figure}[t]
    \centering
    \includegraphics[width=.48\textwidth]{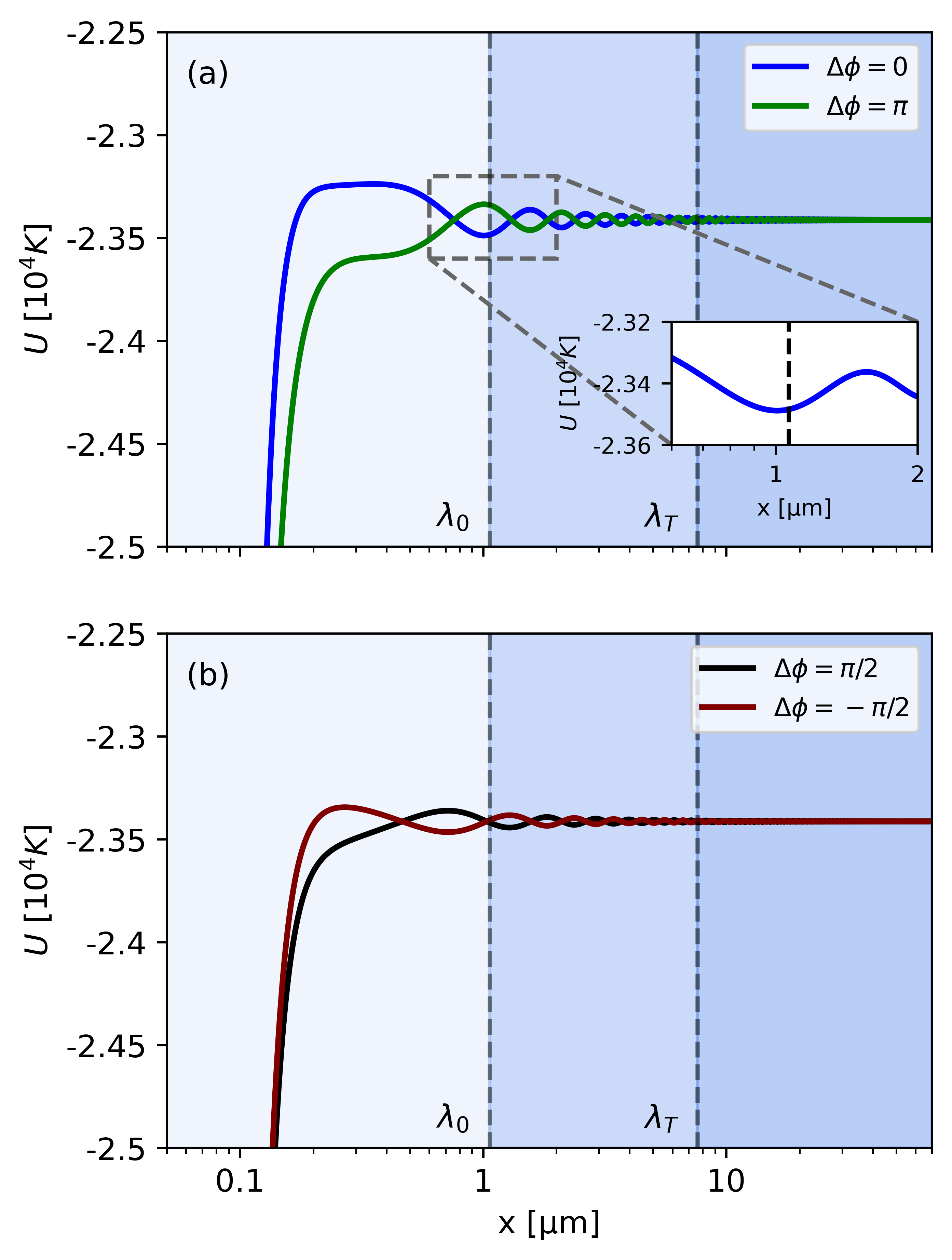}
    \caption{\label{fig:tot_pot} (a) Total potential for $\Delta \phi =0$ (green) and $\Delta \phi = \pi$ (blue) at intensity $I=10^{-2}~ \text{W}/\mu \text{m}^2$, $T=300$~K and with both lasers polarized along the y-axis. The spheres have radius $R=100$~ nm. For these relative optical phases the system is conservative. The potential thus represents the joint potential of both spheres.The inset shows the deepest potential well around $r=\lambda_0$. The plateau of the potential at large distances is formed by the flat tweezer potential. (b) Total potential seen by sphere A (maroon) and sphere B (black) at $\Delta\phi=\pi/2$. Since they now exhibits non-conservative forces, there is no joint potential, but sphere A will see an optical binding potential with $\Delta\phi=\pi/2$ and sphere B with $\Delta\phi=-\pi/2$. The length scales $\lambda_0,\lambda_{t_2}$ and $\lambda_T$ are defined in Sec. \ref{sec:scaling regimes}.}
\end{figure}

\subsubsection{Optical binding potential}
\label{sec:OB_approx}
Under the assumption that the drive is comprised of a single mode, we can use the single-mode expression for optical binding potential in Eq. (\ref{eq:single_mode_OB}). The  characteristic length scale of the optical binding potential is determined by  the tweezer wavelength $\lambda_0$. If we also assume a flat tweezer profile, then, the potential in the near-field regime ($r\ll\lambda_0$) is: 

\begin{align}
\label{eq:OB_near}
    U^\mr{A/B}_\mr{OB}(\mb{r}_A, \mb{r}_B)
    \approx4\pi\frac{\sqrt{I_AI_B}}{c}R^3\bkt{\frac{R}{r}}^3\Tilde{\alpha}(\omega_0)^2\text{cos}(\Delta \phi)
\end{align}
where $I_{A/B}$ is the intensity of the tweezers A and B, $\Delta\phi$ is the relative optical phase between the two tweezers and $\omega_0$ is the drive frequency.

\begin{figure}[t]
    \centering
    \includegraphics[width=0.48\textwidth]{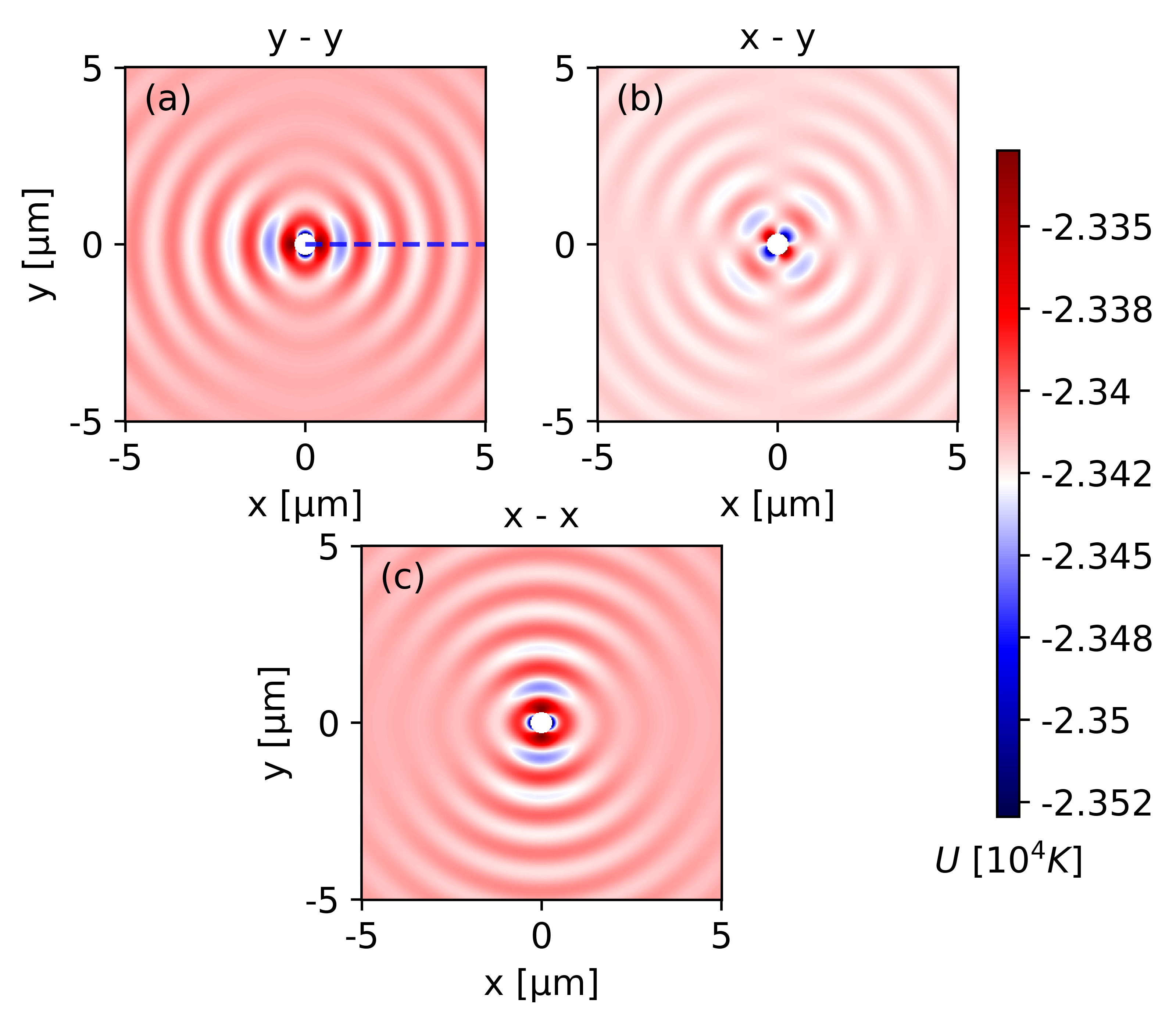}
    \caption{\label{fig:polarization} The total potential for different polarizations of the tweezer: (a) yy-polarization, (b) xy-polarization and (c) xx-polarization, in the $z=0$ plane.  The dashed blue line indicates the potential plotted in Fig. \ref{fig:tot_pot} (a). Regarding interactions along the x-axis, as shown in Fig \ref{fig:tot_pot}, it can be seen that going from yy-polarization to xx-polarization enables one to turn off the optical binding interaction between the spheres. This is due to the radiation pattern of a dipole, which radiates perpendicular to the induced polarization. We assume the intensity to be $I=10^{-2}~ \text{W}/\mu \text{m}^2$ and $\Delta \phi =0$. In the intermediate arrangement, at xy-polarization, there is weak interaction along both the x and the y-axis, due to the interference of the two dipole fields.}
\end{figure}

In the far-field $(r\gg\lambda_0)$ we get:
\begin{align}
\label{eq:OB_far}
    &U^\mr{A/B}_\mr{OB}(\mb{r}_A, \mb{r}_B)\non\\
    &\approx -4\pi\frac{\sqrt{I_AI_B}}{c}R^3(k_0R)^2\bkt{\frac{R}{r}}\Tilde{\alpha}(\omega_0)^2\text{cos}(k_0r\pm\Delta \phi),
\end{align}
which scales with the square root of the radiation pressure $\sim \sqrt{I/c}$ on either sphere. It is worth pointing out, that in the radiative regime when $r\approx \lambda_0$, the potential scales as $\sim \text{sin}(k_0r\pm\Delta \phi)/r^2$, meaning that it exhibits a phase shift of $\pi/2$ compared to the near-field and far-field regimes. 

The two spheres, A and B, can see different optical binding potentials for $\Delta\phi \neq 0$ or $\pi$, resulting in non-conservative forces between the trapped nanospheres. As illustrated in the next section, the difference between potentials is the largest for $\Delta \phi = \pi/2$ and vanishes for $ \Delta \phi = 0 $ or $ \pi $. Such non-conservative optical binding forces were experimentally observed by Rieser et al. in \cite{Rieser22}. 

Utilizing the different scaling behavior of the interparticle potentials along with the fact that the optical binding  and trapping potentials can be readily tuned via the tweezer field one can realize various sphere-sphere potential landscapes. In the next section we will investigate how this can be used to produce a bound state potential of the two-nanospheres.

\subsection{Bound state of nanospheres}
\label{Sec:Bound State}

\begin{figure*}[t]
    \centering
    \includegraphics[width=\textwidth]{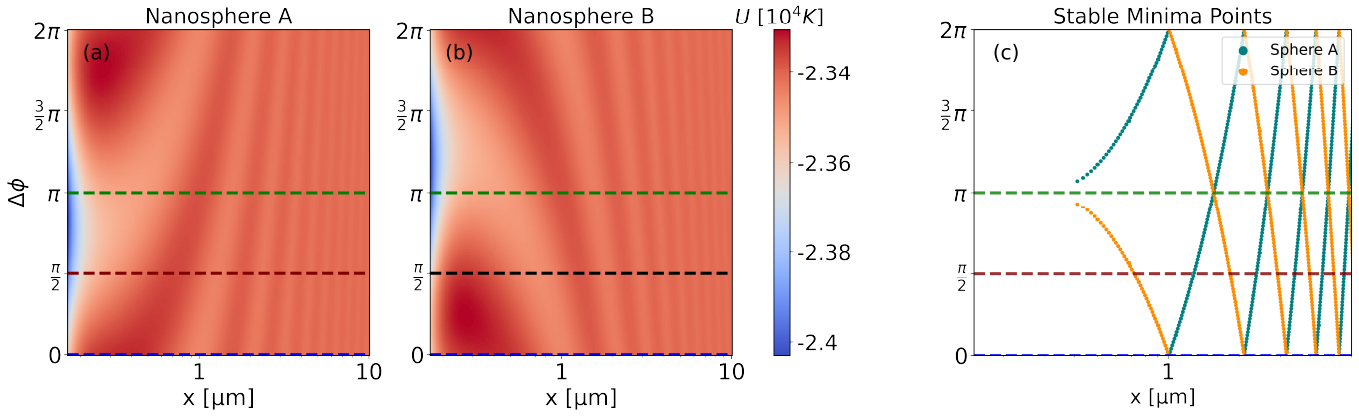}
    \caption{\label{fig:pot_phase}The total potential as a function of relative optical phase $\Delta\phi$ at intensity $I=10^{-2} ~ \mr{W}/\mu \mr{m}^2$, $T=300$~K and with both lasers polarized along the y-axis by sphere (a) A and (b) B. The spheres have radius $R=100$~ nm. The dashed coloured lines coincide with he potentials plotted in Fig. \ref{fig:tot_pot}. (c) Stability analysis along the x-axis of the two nanosphere system. The dotted lines show the minima of the potential of sphere A (teal) and sphere B (orange) as presented in Fig. \ref{fig:pot_phase}. The dashed lines correspond to the potentials in Fig. \ref{fig:tot_pot}. Due to the non-conservative nature of the optical binding interaction, stable bound states along the x-axis can only be found at $\Delta\phi=0$ and $\Delta\phi=\pi$, since only then both spheres can be at a potential minimum simultaneously.}
\end{figure*}

The interplay between the different radiative potentials gives rise to a tunable potential landscape. While the analogue of the Earnshaw's theorem for fluctuation-induced forces fundamentally constrains the possibility of creating stable equilibria for objects interacting purely via quantum fluctuations \cite{Rahi10}, driven systems  can overcome this limitation.
The drive thus allows one to create repulsive and tunable  optical binding potentials, which can be modified by changing the optical tweezers intensity $I$, relative optical phase $\Delta\phi$ and polarizations. 

Fig.~\ref{fig:tot_pot} shows the interparticle potential at $ T = 300$~K for a drive intensity $I=10^{-2}~ \text{W}/\mu \text{m}^2$, with both tweezer fields polarized along the y axis for four different relative optical phases. Fig.~\ref{fig:tot_pot}~(a) shows the potentials for $\Delta\phi=0$ (green) and $\Delta \phi=\pi$ (blue). For these two relative optical phases the forces are purely conservative and a joint potential of both spheres can be written down. It can be seen from the inset in (a) that for $\Delta\phi=0$, there is a $\approx 200$~K deep well around $x=\lambda_0$. In the flat tweezer approximation, this potential well -- and other wells at a greater distance -- are created by the oscillatory nature of the optical binding potential. Their depth therefore increases linearly with the tweezer intensity, which is a readily tunable parameter in the experiment. Fig. \ref{fig:tot_pot}~(b) shows the potential seen by either sphere at $\Delta\phi=\pi/2$. In that case the system is no longer conservative and each sphere sees a different potential. Consequently, for the values of the sphere-sphere separation $x$ where sphere A sees a potential minimum, sphere B sees a maximum. Such a scenario will lead to an unstable configuration of the two spheres, wherein they would eventually attract each other in the near-field and stick together.

Fig. \ref{fig:polarization} shows the total potential for three different polarization configurations of the two tweezer fields applied at $ \mb{r}_A $ and $ \mb{r}_B$ (yy, xy and xx) with  $I=10^{-2} ~ \mr{W}/\mu \mr{m}^2$ and $T=300$~K. It can be seen that the potential reflects the interference of the  dipole radiation pattern from the two nanospheres, thus permitting one to turn off the optical binding interaction between between the spheres. 

In order to study the existence of stable minima in more detail -- taking into account the non-conservative nature of the interaction -- one must consider the potential of either sphere and its dependence on $\Delta \phi$. Fig. \ref{fig:pot_phase}~(a) and (b) present the potentials seen by either sphere as a function of relative optical phase $\Delta \phi$.  In the single mode approximation of the tweezer the optical binding potential of spheres A and B satisfy $U^\mr{A}_\mr{OB}(\Delta\phi)=U^\mr{B}_\mr{OB}(-\Delta\phi)$, such that the potentials are equivalent, but reflected around $\Delta \phi=\pi$. The system is stable if both spheres experience a potential minimum at a distance x. To search for such positions, we numerically find the minima of both potentials in Fig.~\ref{fig:pot_phase} (a) and (b); the results are shown in Fig.~\ref{fig:pot_phase}~(c). The dotted lines in teal and orange represent the minima of the potential of sphere A and sphere B respectively. 
 Since stability is only possible whenever the two dotted lines cross -- indicating that both spheres are at a potential minimum -- it can be seen that bound states only exist at $\Delta\phi=0$ and $\Delta\phi=\pi$, i.e. when there are no non-conservative forces acting on the spheres.

\section{Summary}
\label{Sec:summary}
In this work we have presented a description of the radiative forces between two dielectric nanospheres interacting via the quantum and thermal fluctuations of the EM field, as well as an external drive in a general quantum state. We analyze the interaction between the total field, comprising the field fluctuations and externally applied fields, and the dipole moments induced in the nanospheres by summing over the light scattering processes up to second order in the particle polarizabilities. Considering the two nanospheres to be trapped by a laser, we demonstrate that the tweezer intensity,  polarizations and the relative optical phase between the tweezers provides us with the means to control the interparticle potential. 

\section{Discussion and Outlook}
\label{Discussion}

Analyzing the interaction potential between the spheres to second-order in the particle polarizabilities,  we recover the known expressions for the trapping, optical binding and thermal Casimir-Polder potentials when the external field is in a coherent state and the fluctuation field in a thermal state (Sec.~\ref{Sec:Coh}). For a coherent state of the external field, the two spheres see a mutually bound trap potential as deep as $\approx 200 $~K at a separation of $ x\approx1~\mu$m for a tweezer intensity of $ I = 10^{-2}$~W/$\mu$m$^2$ and phase $\Delta\phi=0$ (Fig.~\ref{fig:tot_pot}).  The relative optical phase between the tweezers permits the creation of non-conservative and non-reciprocal forces. Utilizing the tweezer polarizations, we can selectively turn-off the optical binding interaction between the spheres. 

Tailoring quantum  statistics of light can have exciting applications in many optical phenomena ~\cite{Romero-Isart_Squeezed,Aasi2013, Treps2002}. We show that an external field in a single-mode squeezed vacuum state creates a potential similar to the optical trapping and optical binding potentials,  remarkably in the absence of any coherent field amplitude (Sec.~\ref{Sec:Sq}). The  squeezing required for creating a strong enough potential to trap a nanosphere of radius $ \sim 100 $~nm, however, is substantially large ($\sim 240$ dB). For the external field in a cat state, the resulting potential differs from the coherent state by a factor of $ \sim (1 - \cos \theta)$, where  $\theta$ refers to the phase between the superposed coherent states (Sec.~\ref{Sec:Cat}).

Understanding and controlling the radiative interactions between optically trapped nanospheres is crucial for future experimental studies of  macroscopic quantum systems. Such levitated nanoparticles in the quantum regime provide an ideal testbed for exploring gravitational interactions between quantum systems, paving the way for exploring the potential role of gravity in engendering decoherence and bringing about the quantum-to-classical transition. In such systems it becomes imperative to control the electromagnetic interactions that are fundamentally far stronger than gravity\footnote{For reference, the ratio of electric to gravitational forces between two electrons is $e^2/(4\pi \epsilon_0 G m_e^2)\sim 10^{42}$, where $m_e$ is the electron mass.},  in order to delineate gravitational effects. It will be pertinent to extend the present work to include finite-size effects such that one can prepare bound states of the nanospheres at shorter distances or larger nanospheres and analyze gravitational interactions between the particles in such a regime.

 Having a full description of the radiative  interactions will allow one to engineer the interparticle potential in more detail, including other interactions such as Coulomb or magnetic interactions. As we have  illustrated, one can realize mutually trapped bound states of nanospheres, with the potential depths comparable to  $\approx 200 $~K. Such trap depths can be readily increased by increasing the tweezer intensity, enabling one to realize such bound states of nanoparticles at room temperatures. Such systems can be excellent sensors of external electromagnetic fields, forces and have been used in developing probes for detection of new physics beyond the standard model.

Furthermore, it would be interesting to extend the present results to consider the spheres in a  delocalized superposition of their center-of-mass positions, or with their centers-of-mass being prepared in entangled states. In such a scenario one can ask how  quantum fluctuations provide a fundamental limit to decoherence of macroscopic quantum systems. On the other hand, it would be relevant to consider the effect of the various quantum states of center-of-mass of the two nanospheres on their individual and mutual quantum fluctuation effects -- such as fluctuation forces and decoherence.

\section{Acknowledgements}
 We are grateful to Pierre Meystre and Paulo A. Maia Neto for helpful discussions and feedback on the manuscript. We thank Markus Aspelmeyer for his support.  K.S. acknowledges support from the National Science Foundation under Grant No. PHY-2309341, and by the John Templeton Foundation under Award No. 62422. C.J., U.D., and K.S. acknowledge support by the John Templeton Foundation under Award No. 63033. U.D. acknowledges support from the FWF (Austrian Science Fund, Project No. I 5111-N). This work was supported by  NSF Grant No. PHY-1748958.

\appendix

\section{Green's Tensor}
\label{app_A}

The Green's tensor is the solution to the inhomogeneous Helmholtz equation:

\begin{align}
    \sbkt{\nabla\times \frac{1}{\mu(\mb{r},\omega)}\nabla\times-\frac{\omega^2}{c^2}\epsilon(\mb{r},\omega)}G(\mb{r},\mb{r}',\omega)=\delta(\mb{r}-\mr{r}'), 
\end{align}
where $\mu(\mb{r},\omega)$, $\epsilon(\mb{r},\omega)$ are the spatially dependent relative permeability and relative permittivity respectively that include the presence of any media.

The electric field sourced by a noise current density $ \mb{j}(\mb{r}', \omega)$ for the given boundary conditions is given by:

\begin{align}
    \mb{E}(\mb{r},\omega)=i\mu_0\omega\int d^3\mb{r}\,G(\mb{r},\mb{r}',\omega)\cdot \mb{j}(\mb{r}',\omega)
\end{align}

The  noise current density can be related to the noise polarization and magnetization as $\hat{\mb{j}}_N=-i\omega \hat{\mb{P}}_N +\nabla\times\hat{\mb{M}}_N$, which describe the source of the fluctuation field~\cite{Buhmann1,Buhmann2}. Thus, the electric field is:

\begin{align}
\label{eq:app_A_Ef}
    \hat{\mb{E}}_\mr{f}(\mb{r},\omega)=i\mu_0\omega\int d^3\mb{r} \, G(\mb{r},\mb{r}',\omega)\cdot \bkt{-i\omega \hat{\mb{P}}_N +\nabla\times\hat{\mb{M}}_N}
\end{align}

Defining the bosonic operators $\hat{\mb{f}}_\lambda(\mb{r},\omega)$ associated with the noise polarization $(\lambda = e) $ and magnetization  $(\lambda = m)$  as:
\begin{align}
    \hat{\mb{P}}_N(\mb{r},\omega)&=i\sqrt{\frac{\hbar\epsilon_0}{\pi}\im \ \epsilon(\mb{r},\omega)}\hat{\mb{f}}_e(\mb{r},\omega)\\
    \hat{\mb{M}}_N(\mb{r},\omega)&=\sqrt{\frac{\hbar}{\pi\mu_0}\frac{\im \ \mu(\mb{r},\omega)}{\abs{\mu(\mb{r},\omega)}}}\hat{\mb{f}}_m(\mb{r},\omega)
\end{align}
 and substituting into Eq.~\eqref{eq:app_A_Ef} permits us to define modified Green's tensors:
 
 \begin{align}
     G_e(\mb{r},\mb{r}',\omega)=&i\frac{\omega^2}{c^2}\sqrt{\frac{\hbar\epsilon_0}{\pi}\im \ \epsilon(\mb{r},\omega)}G(\mb{r},\mb{r}',\omega)\\
     G_m(\mb{r},\mb{r}',\omega)=&i\frac{\omega}{c}\sqrt{\frac{\hbar}{\pi\mu_0}\frac{\im \ \mu(\mb{r},\omega)}{\abs{\mu(\mb{r},\omega)}}}\sbkt{\nabla'\times G(\mb{r}',\mb{r},\omega}^T
 \end{align}
 
 This, in turn, we can use to write the fluctuation field as we have done in Eq.~\eqref{eq:E0} as:
 
\eqn{
\hat{\mb{E}}_\mr{f}(\mb{r}_i)=\sum_{\lambda=e,m}\int d^3\mb{r} \int_0^\infty d\omega\, G^{\lambda}(\mb{r}_i, \mb{r},\omega)\cdot\hat{\mb{f}}_\lambda(\mr{r},\omega) +\mr{H. c.}
}

\section{Radiative Forces in optical tweezers}
\label{App_B}

In Sec. \ref{Sec:Coh} we have calculated the potential seen by two nanospheres when the external field is in a coherent state. Here we elaborate on those results and show how they connect to more commonly known aspects of optical trapping and optical binding of nanospheres. To make this comparison we will first map the external field in a coherent state onto the tweezer field and the scattered tweezer field as follows. We define the \textit{complex} fields as: 

\begin{align}
    &\mb E_{\mr{tw}}(\mb r_A)=\bra{\beta}\hat{\mb E}^{(0)}_\mr{ex}(\mb r_A)\ket{\beta}= \non\\
    &= \frac{1}{2}\sum_{\sigma}\int d^3\mb{k}\,\mb{\Phi}_\sigma(\mb{r}_A,\mb{k},\omega)\beta_\sigma(\mb{k})e^{-i\omega t}
\end{align}
and similarly the dipole moment induced by the tweezer field as: 

\begin{align}
    &\mb P_{\mr{tw}}(\mb r_A)=\bra{\beta}\hat{\mb P}^{(1)}_\mr{ex}(\mb r_A)\ket{\beta}= \non\\
    &= \frac{1}{2}\sum_{\sigma}\int d^3\mb{k}\,\alpha(\omega)\mb{\Phi}_\sigma(\mb{r}_A,\mb{k},\omega)\beta_\sigma(\mb{k})e^{-i\omega t}.
\end{align}

The scattered field, namely the fields scattered scattered off sphere A or sphere B and seen by sphere A, we can thus write as: 

\begin{align}
    &\mb E^\mr{A/B}_{\mr{sc}}(\mb r_A)=\bra{\beta}\hat{\mb E}^{(1)}_\mr{ex}(\mb r_A)\ket{\beta}= \non\\
    &= \frac{1}{2}\sum_{\sigma}\int d^3\mb{k}\,G(\mb r_A,\mb r_{A/B},\omega)\mb{\Phi}_\sigma(\mb{r}_{A/B},\mb{k},\omega)\beta_\sigma(\mb{k})e^{-i\omega t}
\end{align}

Where the superscript in $\mb E^\mr{A/B}_{\mr{sc}}(\mb r_A)$ denotes whether the tweezer field is scattered off sphere A or sphere B. Once again, this implies that the dipole moment induced by the scattered tweezer field is: 

\eqn{
    \mb P^\mr{A/B}_{\mr{sc}}(\mb r_A)=\bra{\beta}\hat{\mb P}^{(2)}_\mr{ex}(\mb r_A)\ket{\beta}= \frac{1}{2}\sum_{\sigma}\int d^3\mb{k}\,\alpha(\omega)\non\\
G(\mb r_A,\mb r_{A/B},\omega)\mb{\Phi}_\sigma(\mb{r}_{A/B},\mb{k},\omega)\beta_\sigma(\mb{k})e^{-i\omega t}
}

Using these definitions allows us to write the coherent state potentials:

 \eqn{\label{eq:U1coh}
    U_{A,\mr{coh}}^ {(1)}(\mb{r}_A)=&-\sum_{\sigma\sigma'}\int d^3\mb{k}\int d^3\mb{k'}\,\alpha(\omega)\non\\
    &    \re \ \cbkt{\mb{\Phi}_{\sigma'}^{A\dagger}\cdot \mb{\Phi}_{\sigma}^A\beta_{\sigma'}^*(\mb{k}')\beta_\sigma(\mb{k})e^{-i(\omega-\omega')t}}}
 
and 
\begin{widetext}
    \eqn{
    \label{eq:U2coh}
    &U_{A,\mr{ex}}^{(2)}(\mb r_A ,\mb r_B)=\non\\
    &-\int d^3\mb{k}\int d^3\mb{k}'\,\mu_0\omega^2\alpha(\omega)\alpha(\omega')\re \ \cbkt{\sum_{\sigma\sigma'}\bkt{\mb{\Phi}_{\sigma'}^{A\dagger}\cdot G^\dagger_{AA}(\omega')\cdot \mb{\Phi}^A_{\sigma}+\mb{\Phi}_{\sigma'}^{A\dagger}\cdot G_{AA}(\omega)\cdot \mb{\Phi}^A_{\sigma}}\beta_\sigma(\mb k)^*\beta_\sigma(\mb k') e^{-i(\omega-\omega')t}}\non\\
     &-\int d^3\mb{k}\int d^3\mb{k}'\,\mu_0\omega^2\alpha(\omega)\alpha(\omega')\re \ \cbkt{\sum_{\sigma\sigma'}\bkt{\mb{\Phi}_{\sigma'}^{A\dagger}\cdot G^\dagger_{AB}(\omega')\cdot \mb{\Phi}^A_{\sigma}+\mb{\Phi}_{\sigma'}^{A\dagger}\cdot G_{AB}(\omega)\cdot \mb{\Phi}^A_{\sigma}}\beta_\sigma(\mb k)^*\beta_\sigma(\mb k') e^{-i(\omega-\omega')t}}
}
\end{widetext}

 as

\begin{align}
    U_{A,\mr{coh}}^ {(1)}(\mb{r}_A)=-\frac{1}{4}\mb P^*_{\mr{tw}}(\mb r_A)\cdot\mb E_{\mr{tw}}(\mb r_A)
\end{align}
and
\begin{align}
    U_{A,\mr{ex}}^{(2)}(\mb r_A ,\mb r_B)=&-\frac{1}{2}\sum_{i=A,B}\re \cbkt{\mb P^\mr{i}_{\mr{sc}}(\mb r_A) \cdot\mb E_{\mr{tw}}(\mb r_A)}\non\\
    &-\frac{1}{2}\sum_{i=A,B}\re \cbkt{\mb E^\mr{i}_{\mr{sc}}(\mb r_A) \cdot\mb P_{\mr{tw}}(\mb r_A) }
\end{align}
If we assume that the tweezer comprises only one single mode $\omega_0$ then we can write these potentials in a more familiar form, showing that we correctly recover the tweezer and the optical binding potential \cite{Pavel2010, Burns1989}.

\begin{align}
\label{eq:Btweezer}
    U_{A,\mr{coh}}^ {(1)}(\mb{r}_A)=-\frac{1}{4}\alpha(\omega_0)\abs{\mb E_{\mr{tw}}(\mb r_A)}^2
\end{align}
and
\begin{align}
\label{eq:BOB}
    &U_{A,\mr{ex}}^{(2)}(\mb r_A ,\mb r_B)=\non\\
    &-\frac{1}{2}\mu_0\omega_0^2\alpha(\omega_0)^2\re \ \cbkt{\mb E^*_{\mr{tw}}(\mb r_A)\cdot G(\mb r_A,\mb r_A,\omega_0)\cdot \mb E_{\mr{tw}}(\mb r_A)}\non\\
    &-\frac{1}{2}\mu_0\omega_0^2\alpha(\omega_0)^2\re \ \cbkt{\mb E^*_{\mr{tw}}(\mb r_A)\cdot G(\mb r_A,\mb r_B,\omega_0)\cdot \mb E_{\mr{tw}}(\mb r_B)},
\end{align}
where Eq. \eqref{eq:Btweezer} is the trapping potential created by the tweezer and Eq. \eqref{eq:BOB} is the optical binding potential seen by sphere A.

\section{Free space Green's tensor and approximate potentials}

The free space Green's tensor between  points $\mb{r}_1$ and $\mb{r}_2$ is given by:

\eqn{
\label{greenfree}
{G}_{\mr{free}} \bkt{\mb{r}_1,\mb{r}_2, \omega} =
\frac{e^{ikr}}{4\pi k^2 r^3}\cbkt{f(kr)\mathbb{1}-h(kr)\mb{e}_r\otimes\mb{e}_r}
}
where $f\bkt{x}\equiv1-ix-x^2$, 
$h\bkt{x}\equiv3-3ix-x^2$, $r = \abs{\mb{r}_1- \mb{r}_2}$, and $\mb{e}_r=\mb{r}/r$ is the unit vector connecting the positions of the two spheres.
\subsection{Thermal CP-potential}
Utilizing this form of the Green's tensor, we will now briefly outline the derivations of the approximations in Sec. \ref{sec:scaling regimes}. We begin by writing the thermal CP-potential as:
\begin{widetext}
\eqn{
\label{eq:intermediate_integral}
U(\mb{r}_A, \mb{r}_B)=-\frac{\hbar c}{16\pi^3 \epsilon_0^2  r^7}\int_0^\infty dx \alpha(xc/r)^2\im \ \cbkt{e^{-2ix}\sbkt{3+6ix+5\bkt{ix}^2+2\bkt{ix}^3+\bkt{ix}^4}}\text{coth}\bkt{\frac{\hbar c x}{2k_BTr}}
}
\end{widetext}

where we have used that $\omega/k=c$, $k=\tilde{k}k_{t_2}$ and subsequently set $\tilde{k}k_{t_2} r=x$. Here $k_{t_2} = 2\pi/\lambda_{t_2}$ is the wavenumber corresponding to the dominant transition wavelength  $\lambda_{t_2}$ of silica as defined in Sec. \ref{sec:thermal_CP_approx}.  In the regime where ($r\gg\lambda_{t_2}$)  the polarisability is well approximated by the static polarisability $\alpha(0)$. Utilizing this approximation, we can see that within this regime there is now another regime defined by the thermal length scale $\lambda_T  \equiv \frac{\hbar c}{k_B T}$. We thus distinguish between the two cases $\lambda_T\gg r$ (far-field regime) and $\lambda_T \ll r$ (intermediate regime). We discuss these two cases below:
\begin{enumerate}
    \item {For $\lambda_T\gg r$, we can set $\text{coth}\bkt{\frac{\lambda_T x}{2r}}\approx1$, which allows us to Wick rotate the integral in Eq. \ref{eq:intermediate_integral} and evaluate it to get:

\eqn{
    U_\mr{CP}^{\text{ts}}(\mb{r}_A, \mb{r}_B)\approx-\frac{\hbar c}{16\pi^3\epsilon_0^2r^7}\alpha(0)^2\frac{23}{4},
}
 exhibiting a $r^{-7}$ scaling. Using the dimensionless quantities defined in Sec. \ref{sec:thermal_CP_approx}, we recover Eq. \ref{eq:thermal_CP_intermediate} in the main text.
}
\item{ For $\lambda_T\ll r$, we extend the integral down to $-\infty$ by splitting up the imaginary part and using the fact that $\text{coth}(x)$ is an odd function. We note that the distance between the poles of the integrand in Eq. \ref{eq:intermediate_integral}, which are given by the Matsubara frequencies $\omega_n= \frac{2\pi k_B T}{\hbar}n$, is bigger than $c/r$, which is the decay length of the exponential in the integrand. This means that because of the decaying exponential, only the first residue at the origin will contribute to the sum. This gives: 
 
\begin{align}
     U_\mr{CP}^{\text{ts}}(\mb{r}_A, \mb{r}_B)\approx--\frac{2k_B T }{16\pi^3\epsilon_0^2r^6}\alpha(0)^2 3\pi,
\end{align} which, after substitution of the dimensionless quantities, yields  Eq. \ref{eq:thermal_CP_far}. Thus the potential returns to a $r^{-6}$ scaling \cite{Buhmann2}. }
\end{enumerate}
 
 In the near-field ($r\ll\lambda_{t_2}$) regime we can approximate the potential as:

\eqn{
    U(\mb{r}_A, \mb{r}_B)\approx-\frac{3\hbar}{16\pi^3 \epsilon_0^2 r^6i}\int_{-\infty}^\infty d\omega \alpha(\omega)^2\text{coth}\bkt{\frac{\hbar\omega}{2k_BT}}.
}
Where we have again extended the integral by breaking up the imaginary part before neglecting the last higher order terms. The integral can be simplified by writing $\text{coth}\bkt{\frac{\hbar\omega}{2k_BT}}=2n(\omega)+1$ and using the same residues as before and a semi-circle contour around the upper half plane. We therefore get:

\eqn{
     &U_\mr{CP}^{\text{ts}}(\mb{r}_A, \mb{r}_B)\approx \non\\
     &-\frac{\hbar}{16\pi^3 \epsilon_0^2 r^6}\cbkt{2\pi \frac{2k_BT}{\hbar}\sum_j \alpha(i\xi_j)^2+\int_0^{\infty}d\omega\alpha(\omega)^2}
}
The first term represents the near field approximation to the thermal part of the potential whereas the second term represents the approximation to the ground state potential. Once again using the dimensionless quantities in Sec. \ref{sec:scaling regimes} we end up with Eq. \ref{eq:thermal_CP_near}. 

\subsection{Optical Binding Potential}

Next we outline the derivation of the approximation to the optical binding potential as given in Sec. \ref{sec:OB_approx}. We work under two assumtptions. Firstly, that the two spheres are confined to move along the x-axis and secondly, that the external field is y-polarized. The coordinate system is defined as in Fig. \ref{Fig:Sch}. Under these assumptions, we can write the OB-potential as given in Eq. \ref{eq:free_space_OB} is: 

\eqn{
&U_{OB}^\mr{A/B}(\mb{r}_A, \mb{r}_B)=\frac{4\pi R^6 \sqrt{I_AI_B}\Tilde{\alpha}(\omega)^2}{c r^3}\non \\
&\cbkt{\text{cos}(kr-\Delta\phi)+\text{sin}(kr-\Delta\phi)kr-\text{cos}(kr-\Delta)(kr)^2},
}
where the dimensionless polarizability $\Tilde{\alpha}$ is defined in Sec. \ref{sec:thermal_CP_approx}. In the near-field regime $(r\ll \lambda_0)$ only the lowest order term in $kr$ contributes, which means that the potential is
\eqn{
    U^\mr{A/B}_\mr{OB}(\mb{r}_A, \mb{r}_B)
    \approx4\pi\frac{\sqrt{I_AI_B}}{c}R^3\bkt{\frac{R}{r}}^3\Tilde{\alpha}(\omega_0)^2\text{cos}(\Delta \phi)
}
as in Eq. \ref{eq:OB_near}. Contrastingly, in the far-field regime $(r\gg \lambda_0)$ only the highest order term in $kr$ contributes, giving us the approximate potential in Eq. \ref{eq:OB_far}.

\bibliography{nanosphere}

\def\bibsection{\section*{}}\providecommand{\noopsort}[1]{}\providecommand{\singleletter}[1]{#1}
\begin{thebibliography}{43}%
\makeatletter
\providecommand \@ifxundefined [1]{%
 \@ifx{#1\undefined}
}%
\providecommand \@ifnum [1]{%
 \ifnum #1\expandafter \@firstoftwo
 \else \expandafter \@secondoftwo
 \fi
}%
\providecommand \@ifx [1]{%
 \ifx #1\expandafter \@firstoftwo
 \else \expandafter \@secondoftwo
 \fi
}%
\providecommand \natexlab [1]{#1}%
\providecommand \enquote  [1]{``#1''}%
\providecommand \bibnamefont  [1]{#1}%
\providecommand \bibfnamefont [1]{#1}%
\providecommand \citenamefont [1]{#1}%
\providecommand \href@noop [0]{\@secondoftwo}%
\providecommand \href [0]{\begingroup \@sanitize@url \@href}%
\providecommand \@href[1]{\@@startlink{#1}\@@href}%
\providecommand \@@href[1]{\endgroup#1\@@endlink}%
\providecommand \@sanitize@url [0]{\catcode `\\12\catcode `\$12\catcode `\&12\catcode `\#12\catcode `\^12\catcode `\_12\catcode `\%12\relax}%
\providecommand \@@startlink[1]{}%
\providecommand \@@endlink[0]{}%
\providecommand \url  [0]{\begingroup\@sanitize@url \@url }%
\providecommand \@url [1]{\endgroup\@href {#1}{\urlprefix }}%
\providecommand \urlprefix  [0]{URL }%
\providecommand \Eprint [0]{\href }%
\providecommand \doibase [0]{https://doi.org/}%
\providecommand \selectlanguage [0]{\@gobble}%
\providecommand \bibinfo  [0]{\@secondoftwo}%
\providecommand \bibfield  [0]{\@secondoftwo}%
\providecommand \translation [1]{[#1]}%
\providecommand \BibitemOpen [0]{}%
\providecommand \bibitemStop [0]{}%
\providecommand \bibitemNoStop [0]{.\EOS\space}%
\providecommand \EOS [0]{\spacefactor3000\relax}%
\providecommand \BibitemShut  [1]{\csname bibitem#1\endcsname}%
\let\auto@bib@innerbib\@empty
\bibitem [{\citenamefont {Zurek}(1991)}]{Zurek91}%
  \BibitemOpen
  \bibfield  {author} {\bibinfo {author} {\bibfnamefont {W.~H.}\ \bibnamefont {Zurek}},\ }\bibfield  {title} {\bibinfo {title} {Decoherence and the transition from quantum to classical},\ }\href {https://doi.org/10.1063/1.881293} {\bibfield  {journal} {\bibinfo  {journal} {Physics Today}\ }\textbf {\bibinfo {volume} {44}},\ \bibinfo {pages} {36} (\bibinfo {year} {1991})}\BibitemShut {NoStop}%
\bibitem [{\citenamefont {Raimond}\ \emph {et~al.}(2001)\citenamefont {Raimond}, \citenamefont {Brune},\ and\ \citenamefont {Haroche}}]{Raimond2001}%
  \BibitemOpen
  \bibfield  {author} {\bibinfo {author} {\bibfnamefont {J.~M.}\ \bibnamefont {Raimond}}, \bibinfo {author} {\bibfnamefont {M.}~\bibnamefont {Brune}},\ and\ \bibinfo {author} {\bibfnamefont {S.}~\bibnamefont {Haroche}},\ }\bibfield  {title} {\bibinfo {title} {Manipulating quantum entanglement with atoms and photons in a cavity},\ }\href {https://doi.org/10.1103/RevModPhys.73.565} {\bibfield  {journal} {\bibinfo  {journal} {Rev. Mod. Phys.}\ }\textbf {\bibinfo {volume} {73}},\ \bibinfo {pages} {565} (\bibinfo {year} {2001})}\BibitemShut {NoStop}%
\bibitem [{\citenamefont {De~Martini}\ and\ \citenamefont {Sciarrino}(2012)}]{MQPRMP}%
  \BibitemOpen
  \bibfield  {author} {\bibinfo {author} {\bibfnamefont {F.}~\bibnamefont {De~Martini}}\ and\ \bibinfo {author} {\bibfnamefont {F.}~\bibnamefont {Sciarrino}},\ }\bibfield  {title} {\bibinfo {title} {Colloquium: Multiparticle quantum superpositions and the quantum-to-classical transition},\ }\href {https://doi.org/10.1103/RevModPhys.84.1765} {\bibfield  {journal} {\bibinfo  {journal} {Rev. Mod. Phys.}\ }\textbf {\bibinfo {volume} {84}},\ \bibinfo {pages} {1765} (\bibinfo {year} {2012})}\BibitemShut {NoStop}%
\bibitem [{\citenamefont {Arndt}\ and\ \citenamefont {Hornberger}(2014)}]{Arndt2014}%
  \BibitemOpen
  \bibfield  {author} {\bibinfo {author} {\bibfnamefont {M.}~\bibnamefont {Arndt}}\ and\ \bibinfo {author} {\bibfnamefont {K.}~\bibnamefont {Hornberger}},\ }\bibfield  {title} {\bibinfo {title} {Testing the limits of quantum mechanical superpositions},\ }\href {https://doi.org/10.1038/nphys2863} {\bibfield  {journal} {\bibinfo  {journal} {Nature Physics}\ }\textbf {\bibinfo {volume} {10}},\ \bibinfo {pages} {271} (\bibinfo {year} {2014})}\BibitemShut {NoStop}%
\bibitem [{\citenamefont {Del{\'e}glise}\ \emph {et~al.}(2008)\citenamefont {Del{\'e}glise}, \citenamefont {Dotsenko}, \citenamefont {Sayrin}, \citenamefont {Bernu}, \citenamefont {Brune}, \citenamefont {Raimond},\ and\ \citenamefont {Haroche}}]{Deleglise2008}%
  \BibitemOpen
  \bibfield  {author} {\bibinfo {author} {\bibfnamefont {S.}~\bibnamefont {Del{\'e}glise}}, \bibinfo {author} {\bibfnamefont {I.}~\bibnamefont {Dotsenko}}, \bibinfo {author} {\bibfnamefont {C.}~\bibnamefont {Sayrin}}, \bibinfo {author} {\bibfnamefont {J.}~\bibnamefont {Bernu}}, \bibinfo {author} {\bibfnamefont {M.}~\bibnamefont {Brune}}, \bibinfo {author} {\bibfnamefont {J.-M.}\ \bibnamefont {Raimond}},\ and\ \bibinfo {author} {\bibfnamefont {S.}~\bibnamefont {Haroche}},\ }\bibfield  {title} {\bibinfo {title} {Reconstruction of non-classical cavity field states with snapshots of their decoherence},\ }\href {https://doi.org/10.1038/nature07288} {\bibfield  {journal} {\bibinfo  {journal} {Nature}\ }\textbf {\bibinfo {volume} {455}},\ \bibinfo {pages} {510} (\bibinfo {year} {2008})}\BibitemShut {NoStop}%
\bibitem [{\citenamefont {Monroe}\ \emph {et~al.}(1996)\citenamefont {Monroe}, \citenamefont {Meekhof}, \citenamefont {King},\ and\ \citenamefont {Wineland}}]{SCAtom}%
  \BibitemOpen
  \bibfield  {author} {\bibinfo {author} {\bibfnamefont {C.}~\bibnamefont {Monroe}}, \bibinfo {author} {\bibfnamefont {D.~M.}\ \bibnamefont {Meekhof}}, \bibinfo {author} {\bibfnamefont {B.~E.}\ \bibnamefont {King}},\ and\ \bibinfo {author} {\bibfnamefont {D.~J.}\ \bibnamefont {Wineland}},\ }\bibfield  {title} {\bibinfo {title} {A ``{S}chr\" odinger cat'' superposition state of an atom},\ }\href {https://doi.org/10.1126/science.272.5265.1131} {\bibfield  {journal} {\bibinfo  {journal} {Science}\ }\textbf {\bibinfo {volume} {272}},\ \bibinfo {pages} {1131} (\bibinfo {year} {1996})}\BibitemShut {NoStop}%
\bibitem [{\citenamefont {Vlastakis}\ \emph {et~al.}(2013)\citenamefont {Vlastakis}, \citenamefont {Kirchmair}, \citenamefont {Leghtas}, \citenamefont {Nigg}, \citenamefont {Frunzio}, \citenamefont {Girvin}, \citenamefont {Mirrahimi}, \citenamefont {Devoret},\ and\ \citenamefont {Schoelkopf}}]{Vlastakis13}%
  \BibitemOpen
  \bibfield  {author} {\bibinfo {author} {\bibfnamefont {B.}~\bibnamefont {Vlastakis}}, \bibinfo {author} {\bibfnamefont {G.}~\bibnamefont {Kirchmair}}, \bibinfo {author} {\bibfnamefont {Z.}~\bibnamefont {Leghtas}}, \bibinfo {author} {\bibfnamefont {S.~E.}\ \bibnamefont {Nigg}}, \bibinfo {author} {\bibfnamefont {L.}~\bibnamefont {Frunzio}}, \bibinfo {author} {\bibfnamefont {S.~M.}\ \bibnamefont {Girvin}}, \bibinfo {author} {\bibfnamefont {M.}~\bibnamefont {Mirrahimi}}, \bibinfo {author} {\bibfnamefont {M.~H.}\ \bibnamefont {Devoret}},\ and\ \bibinfo {author} {\bibfnamefont {R.~J.}\ \bibnamefont {Schoelkopf}},\ }\bibfield  {title} {\bibinfo {title} {Deterministically encoding quantum information using 100-photon {S}chr\"{o}dinger cat states},\ }\href {https://doi.org/10.1126/science.1243289} {\bibfield  {journal} {\bibinfo  {journal} {Science}\ }\textbf {\bibinfo {volume} {342}},\ \bibinfo {pages} {607} (\bibinfo {year} {2013})}\BibitemShut {NoStop}%
\bibitem [{\citenamefont {Bild}\ \emph {et~al.}(2023)\citenamefont {Bild}, \citenamefont {Fadel}, \citenamefont {Yang}, \citenamefont {von L{\"u}pke}, \citenamefont {Martin}, \citenamefont {Bruno},\ and\ \citenamefont {Chu}}]{Bild2023}%
  \BibitemOpen
  \bibfield  {author} {\bibinfo {author} {\bibfnamefont {M.}~\bibnamefont {Bild}}, \bibinfo {author} {\bibfnamefont {M.}~\bibnamefont {Fadel}}, \bibinfo {author} {\bibfnamefont {Y.}~\bibnamefont {Yang}}, \bibinfo {author} {\bibfnamefont {U.}~\bibnamefont {von L{\"u}pke}}, \bibinfo {author} {\bibfnamefont {P.}~\bibnamefont {Martin}}, \bibinfo {author} {\bibfnamefont {A.}~\bibnamefont {Bruno}},\ and\ \bibinfo {author} {\bibfnamefont {Y.}~\bibnamefont {Chu}},\ }\bibfield  {title} {\bibinfo {title} {Schr{\"o}dinger cat states of a 16-microgram mechanical oscillator},\ }\href {https://doi.org/10.1126/science.adf7553} {\bibfield  {journal} {\bibinfo  {journal} {Science}\ }\textbf {\bibinfo {volume} {380}},\ \bibinfo {pages} {274} (\bibinfo {year} {2023})}\BibitemShut {NoStop}%
\bibitem [{\citenamefont {Fein}\ \emph {et~al.}(2019)\citenamefont {Fein}, \citenamefont {Geyer}, \citenamefont {Zwick}, \citenamefont {Kia{\l}ka}, \citenamefont {Pedalino}, \citenamefont {Mayor}, \citenamefont {Gerlich},\ and\ \citenamefont {Arndt}}]{Fein2019}%
  \BibitemOpen
  \bibfield  {author} {\bibinfo {author} {\bibfnamefont {Y.~Y.}\ \bibnamefont {Fein}}, \bibinfo {author} {\bibfnamefont {P.}~\bibnamefont {Geyer}}, \bibinfo {author} {\bibfnamefont {P.}~\bibnamefont {Zwick}}, \bibinfo {author} {\bibfnamefont {F.}~\bibnamefont {Kia{\l}ka}}, \bibinfo {author} {\bibfnamefont {S.}~\bibnamefont {Pedalino}}, \bibinfo {author} {\bibfnamefont {M.}~\bibnamefont {Mayor}}, \bibinfo {author} {\bibfnamefont {S.}~\bibnamefont {Gerlich}},\ and\ \bibinfo {author} {\bibfnamefont {M.}~\bibnamefont {Arndt}},\ }\bibfield  {title} {\bibinfo {title} {Quantum superposition of molecules beyond 25 kda},\ }\href {https://doi.org/10.1038/s41567-019-0663-9} {\bibfield  {journal} {\bibinfo  {journal} {Nature Physics}\ }\textbf {\bibinfo {volume} {15}},\ \bibinfo {pages} {1242} (\bibinfo {year} {2019})}\BibitemShut {NoStop}%
\bibitem [{\citenamefont {Rossi}\ \emph {et~al.}(2018)\citenamefont {Rossi}, \citenamefont {Mason}, \citenamefont {Chen}, \citenamefont {Tsaturyan},\ and\ \citenamefont {Schliesser}}]{Rossi2018}%
  \BibitemOpen
  \bibfield  {author} {\bibinfo {author} {\bibfnamefont {M.}~\bibnamefont {Rossi}}, \bibinfo {author} {\bibfnamefont {D.}~\bibnamefont {Mason}}, \bibinfo {author} {\bibfnamefont {J.}~\bibnamefont {Chen}}, \bibinfo {author} {\bibfnamefont {Y.}~\bibnamefont {Tsaturyan}},\ and\ \bibinfo {author} {\bibfnamefont {A.}~\bibnamefont {Schliesser}},\ }\bibfield  {title} {\bibinfo {title} {Measurement-based quantum control of mechanical motion},\ }\href {https://doi.org/10.1038/s41586-018-0643-8} {\bibfield  {journal} {\bibinfo  {journal} {Nature}\ }\textbf {\bibinfo {volume} {563}},\ \bibinfo {pages} {53} (\bibinfo {year} {2018})}\BibitemShut {NoStop}%
\bibitem [{\citenamefont {Gonzalez-Ballestero}\ \emph {et~al.}(2021)\citenamefont {Gonzalez-Ballestero}, \citenamefont {Aspelmeyer}, \citenamefont {Novotny}, \citenamefont {Quidant},\ and\ \citenamefont {Romero-Isart}}]{GonzalezBallestero21}%
  \BibitemOpen
  \bibfield  {author} {\bibinfo {author} {\bibfnamefont {C.}~\bibnamefont {Gonzalez-Ballestero}}, \bibinfo {author} {\bibfnamefont {M.}~\bibnamefont {Aspelmeyer}}, \bibinfo {author} {\bibfnamefont {L.}~\bibnamefont {Novotny}}, \bibinfo {author} {\bibfnamefont {R.}~\bibnamefont {Quidant}},\ and\ \bibinfo {author} {\bibfnamefont {O.}~\bibnamefont {Romero-Isart}},\ }\bibfield  {title} {\bibinfo {title} {Levitodynamics: Levitation and control of microscopic objects in vacuum},\ }\href {https://www.science.org/doi/abs/10.1126/science.abg3027} {\bibfield  {journal} {\bibinfo  {journal} {Science}\ }\textbf {\bibinfo {volume} {374}} (\bibinfo {year} {2021})}\BibitemShut {NoStop}%
\bibitem [{\citenamefont {Ashkin}(1970)}]{Ashkin}%
  \BibitemOpen
  \bibfield  {author} {\bibinfo {author} {\bibfnamefont {A.}~\bibnamefont {Ashkin}},\ }\bibfield  {title} {\bibinfo {title} {Acceleration and trapping of particles by radiation pressure},\ }\href {https://doi.org/10.1103/PhysRevLett.24.156} {\bibfield  {journal} {\bibinfo  {journal} {Phys. Rev. Lett.}\ }\textbf {\bibinfo {volume} {24}},\ \bibinfo {pages} {156} (\bibinfo {year} {1970})}\BibitemShut {NoStop}%
\bibitem [{\citenamefont {Deli\'{c}}\ \emph {et~al.}(2020)\citenamefont {Deli\'{c}}, \citenamefont {Reisenbauer}, \citenamefont {Dare}, \citenamefont {Grass}, \citenamefont {Vuleti\'{c}}, \citenamefont {Kiesel},\ and\ \citenamefont {Aspelmeyer}}]{Delic20}%
  \BibitemOpen
  \bibfield  {author} {\bibinfo {author} {\bibfnamefont {U.}~\bibnamefont {Deli\'{c}}}, \bibinfo {author} {\bibfnamefont {M.}~\bibnamefont {Reisenbauer}}, \bibinfo {author} {\bibfnamefont {K.}~\bibnamefont {Dare}}, \bibinfo {author} {\bibfnamefont {D.}~\bibnamefont {Grass}}, \bibinfo {author} {\bibfnamefont {V.}~\bibnamefont {Vuleti\'{c}}}, \bibinfo {author} {\bibfnamefont {N.}~\bibnamefont {Kiesel}},\ and\ \bibinfo {author} {\bibfnamefont {M.}~\bibnamefont {Aspelmeyer}},\ }\bibfield  {title} {\bibinfo {title} {Cooling of a levitated nanoparticle to the motional quantum ground state},\ }\href {https://doi.org/10.1126/science.aba3993} {\bibfield  {journal} {\bibinfo  {journal} {Science}\ }\textbf {\bibinfo {volume} {367}},\ \bibinfo {pages} {892} (\bibinfo {year} {2020})}\BibitemShut {NoStop}%
\bibitem [{\citenamefont {Ranfagni}\ \emph {et~al.}(2022)\citenamefont {Ranfagni}, \citenamefont {B\o{}rkje}, \citenamefont {Marino},\ and\ \citenamefont {Marin}}]{Marin2022}%
  \BibitemOpen
  \bibfield  {author} {\bibinfo {author} {\bibfnamefont {A.}~\bibnamefont {Ranfagni}}, \bibinfo {author} {\bibfnamefont {K.}~\bibnamefont {B\o{}rkje}}, \bibinfo {author} {\bibfnamefont {F.}~\bibnamefont {Marino}},\ and\ \bibinfo {author} {\bibfnamefont {F.}~\bibnamefont {Marin}},\ }\bibfield  {title} {\bibinfo {title} {Two-dimensional quantum motion of a levitated nanosphere},\ }\href {https://doi.org/10.1103/PhysRevResearch.4.033051} {\bibfield  {journal} {\bibinfo  {journal} {Phys. Rev. Res.}\ }\textbf {\bibinfo {volume} {4}},\ \bibinfo {pages} {033051} (\bibinfo {year} {2022})}\BibitemShut {NoStop}%
\bibitem [{\citenamefont {Piotrowski}\ \emph {et~al.}(2023)\citenamefont {Piotrowski}, \citenamefont {Windey}, \citenamefont {Vijayan}, \citenamefont {Gonzalez-Ballestero}, \citenamefont {de~los R{\'i}os~Sommer}, \citenamefont {Meyer}, \citenamefont {Quidant}, \citenamefont {Romero-Isart}, \citenamefont {Reimann},\ and\ \citenamefont {Novotny}}]{Piotrowski2023}%
  \BibitemOpen
  \bibfield  {author} {\bibinfo {author} {\bibfnamefont {J.}~\bibnamefont {Piotrowski}}, \bibinfo {author} {\bibfnamefont {D.}~\bibnamefont {Windey}}, \bibinfo {author} {\bibfnamefont {J.}~\bibnamefont {Vijayan}}, \bibinfo {author} {\bibfnamefont {C.}~\bibnamefont {Gonzalez-Ballestero}}, \bibinfo {author} {\bibfnamefont {A.}~\bibnamefont {de~los R{\'i}os~Sommer}}, \bibinfo {author} {\bibfnamefont {N.}~\bibnamefont {Meyer}}, \bibinfo {author} {\bibfnamefont {R.}~\bibnamefont {Quidant}}, \bibinfo {author} {\bibfnamefont {O.}~\bibnamefont {Romero-Isart}}, \bibinfo {author} {\bibfnamefont {R.}~\bibnamefont {Reimann}},\ and\ \bibinfo {author} {\bibfnamefont {L.}~\bibnamefont {Novotny}},\ }\bibfield  {title} {\bibinfo {title} {Simultaneous ground-state cooling of two mechanical modes of a levitated nanoparticle},\ }\href {https://doi.org/10.1038/s41567-023-01956-1} {\bibfield  {journal} {\bibinfo  {journal} {Nature Physics}\ }\textbf {\bibinfo {volume} {19}},\ \bibinfo {pages} {1009} (\bibinfo {year}
  {2023})}\BibitemShut {NoStop}%
\bibitem [{\citenamefont {Magrini}\ \emph {et~al.}(2021)\citenamefont {Magrini}, \citenamefont {Rosenzweig}, \citenamefont {Bach}, \citenamefont {Deutschmann-Olek}, \citenamefont {Hofer}, \citenamefont {Hong}, \citenamefont {Kiesel}, \citenamefont {Kugi},\ and\ \citenamefont {Aspelmeyer}}]{Magrini2021}%
  \BibitemOpen
  \bibfield  {author} {\bibinfo {author} {\bibfnamefont {L.}~\bibnamefont {Magrini}}, \bibinfo {author} {\bibfnamefont {P.}~\bibnamefont {Rosenzweig}}, \bibinfo {author} {\bibfnamefont {C.}~\bibnamefont {Bach}}, \bibinfo {author} {\bibfnamefont {A.}~\bibnamefont {Deutschmann-Olek}}, \bibinfo {author} {\bibfnamefont {S.~G.}\ \bibnamefont {Hofer}}, \bibinfo {author} {\bibfnamefont {S.}~\bibnamefont {Hong}}, \bibinfo {author} {\bibfnamefont {N.}~\bibnamefont {Kiesel}}, \bibinfo {author} {\bibfnamefont {A.}~\bibnamefont {Kugi}},\ and\ \bibinfo {author} {\bibfnamefont {M.}~\bibnamefont {Aspelmeyer}},\ }\bibfield  {title} {\bibinfo {title} {Real-time optimal quantum control of mechanical motion at room temperature},\ }\href {https://doi.org/10.1038/s41586-021-03602-3} {\bibfield  {journal} {\bibinfo  {journal} {Nature}\ }\textbf {\bibinfo {volume} {595}},\ \bibinfo {pages} {373} (\bibinfo {year} {2021})}\BibitemShut {NoStop}%
\bibitem [{\citenamefont {Tebbenjohanns}\ \emph {et~al.}(2021)\citenamefont {Tebbenjohanns}, \citenamefont {Mattana}, \citenamefont {Rossi}, \citenamefont {Frimmer},\ and\ \citenamefont {Novotny}}]{Tebbenjohanns2021}%
  \BibitemOpen
  \bibfield  {author} {\bibinfo {author} {\bibfnamefont {F.}~\bibnamefont {Tebbenjohanns}}, \bibinfo {author} {\bibfnamefont {M.~L.}\ \bibnamefont {Mattana}}, \bibinfo {author} {\bibfnamefont {M.}~\bibnamefont {Rossi}}, \bibinfo {author} {\bibfnamefont {M.}~\bibnamefont {Frimmer}},\ and\ \bibinfo {author} {\bibfnamefont {L.}~\bibnamefont {Novotny}},\ }\bibfield  {title} {\bibinfo {title} {Quantum control of a nanoparticle optically levitated in cryogenic free space},\ }\href {https://doi.org/10.1038/s41586-021-03617-w} {\bibfield  {journal} {\bibinfo  {journal} {Nature}\ }\textbf {\bibinfo {volume} {595}},\ \bibinfo {pages} {378} (\bibinfo {year} {2021})}\BibitemShut {NoStop}%
\bibitem [{\citenamefont {Kamba}\ \emph {et~al.}(2022)\citenamefont {Kamba}, \citenamefont {Shimizu},\ and\ \citenamefont {Aikawa}}]{Kamba2022}%
  \BibitemOpen
  \bibfield  {author} {\bibinfo {author} {\bibfnamefont {M.}~\bibnamefont {Kamba}}, \bibinfo {author} {\bibfnamefont {R.}~\bibnamefont {Shimizu}},\ and\ \bibinfo {author} {\bibfnamefont {K.}~\bibnamefont {Aikawa}},\ }\bibfield  {title} {\bibinfo {title} {Optical cold damping of neutral nanoparticles near the ground state in an optical lattice},\ }\href {https://doi.org/10.1364/OE.462921} {\bibfield  {journal} {\bibinfo  {journal} {Opt. Express}\ }\textbf {\bibinfo {volume} {30}},\ \bibinfo {pages} {26716} (\bibinfo {year} {2022})}\BibitemShut {NoStop}%
\bibitem [{\citenamefont {Rieser}\ \emph {et~al.}(2022)\citenamefont {Rieser}, \citenamefont {Ciampini}, \citenamefont {Rudolph}, \citenamefont {Kiesel}, \citenamefont {Hornberger}, \citenamefont {Stickler}, \citenamefont {Aspelmeyer},\ and\ \citenamefont {Deli{\'c}}}]{Rieser22}%
  \BibitemOpen
  \bibfield  {author} {\bibinfo {author} {\bibfnamefont {J.}~\bibnamefont {Rieser}}, \bibinfo {author} {\bibfnamefont {M.~A.}\ \bibnamefont {Ciampini}}, \bibinfo {author} {\bibfnamefont {H.}~\bibnamefont {Rudolph}}, \bibinfo {author} {\bibfnamefont {N.}~\bibnamefont {Kiesel}}, \bibinfo {author} {\bibfnamefont {K.}~\bibnamefont {Hornberger}}, \bibinfo {author} {\bibfnamefont {B.~A.}\ \bibnamefont {Stickler}}, \bibinfo {author} {\bibfnamefont {M.}~\bibnamefont {Aspelmeyer}},\ and\ \bibinfo {author} {\bibfnamefont {U.}~\bibnamefont {Deli{\'c}}},\ }\bibfield  {title} {\bibinfo {title} {Tunable light-induced dipole-dipole interaction between optically levitated nanoparticles},\ }\href {https://doi.org/10.1126/science.abp9941} {\bibfield  {journal} {\bibinfo  {journal} {Science}\ }\textbf {\bibinfo {volume} {377}},\ \bibinfo {pages} {987} (\bibinfo {year} {2022})}\BibitemShut {NoStop}%
\bibitem [{\citenamefont {Milonni}(1994)}]{Milonni}%
  \BibitemOpen
  \bibfield  {author} {\bibinfo {author} {\bibfnamefont {P.~W.}\ \bibnamefont {Milonni}},\ }\href {https://books.google.com/books?id=P83vAAAAMAAJ} {\emph {\bibinfo {title} {The Quantum Vacuum: An Introduction to Quantum Electrodynamics}}}\ (\bibinfo  {publisher} {Elsevier Science},\ \bibinfo {year} {1994})\BibitemShut {NoStop}%
\bibitem [{\citenamefont {Rahi}\ \emph {et~al.}(2010)\citenamefont {Rahi}, \citenamefont {Kardar},\ and\ \citenamefont {Emig}}]{Rahi10}%
  \BibitemOpen
  \bibfield  {author} {\bibinfo {author} {\bibfnamefont {S.~J.}\ \bibnamefont {Rahi}}, \bibinfo {author} {\bibfnamefont {M.}~\bibnamefont {Kardar}},\ and\ \bibinfo {author} {\bibfnamefont {T.}~\bibnamefont {Emig}},\ }\bibfield  {title} {\bibinfo {title} {Constraints on stable equilibria with fluctuation-induced (casimir) forces},\ }\href {https://doi.org/10.1103/PhysRevLett.105.070404} {\bibfield  {journal} {\bibinfo  {journal} {Phys. Rev. Lett.}\ }\textbf {\bibinfo {volume} {105}},\ \bibinfo {pages} {070404} (\bibinfo {year} {2010})}\BibitemShut {NoStop}%
\bibitem [{\citenamefont {Agrenius}\ \emph {et~al.}(2023)\citenamefont {Agrenius}, \citenamefont {Gonzalez-Ballestero}, \citenamefont {Maurer},\ and\ \citenamefont {Romero-Isart}}]{Agrenius2023}%
  \BibitemOpen
  \bibfield  {author} {\bibinfo {author} {\bibfnamefont {T.}~\bibnamefont {Agrenius}}, \bibinfo {author} {\bibfnamefont {C.}~\bibnamefont {Gonzalez-Ballestero}}, \bibinfo {author} {\bibfnamefont {P.}~\bibnamefont {Maurer}},\ and\ \bibinfo {author} {\bibfnamefont {O.}~\bibnamefont {Romero-Isart}},\ }\bibfield  {title} {\bibinfo {title} {Interaction between an optically levitated nanoparticle and its thermal image: Internal thermometry via displacement sensing},\ }\href {https://doi.org/10.1103/PhysRevLett.130.093601} {\bibfield  {journal} {\bibinfo  {journal} {Phys. Rev. Lett.}\ }\textbf {\bibinfo {volume} {130}},\ \bibinfo {pages} {093601} (\bibinfo {year} {2023})}\BibitemShut {NoStop}%
\bibitem [{\citenamefont {Milonni}\ and\ \citenamefont {Smith}(1996{\natexlab{a}})}]{MilonniSmith96}%
  \BibitemOpen
  \bibfield  {author} {\bibinfo {author} {\bibfnamefont {P.~W.}\ \bibnamefont {Milonni}}\ and\ \bibinfo {author} {\bibfnamefont {A.}~\bibnamefont {Smith}},\ }\bibfield  {title} {\bibinfo {title} {van der waals dispersion forces in electromagnetic fields},\ }\href {https://doi.org/10.1103/PhysRevA.53.3484} {\bibfield  {journal} {\bibinfo  {journal} {Phys. Rev. A}\ }\textbf {\bibinfo {volume} {53}},\ \bibinfo {pages} {3484} (\bibinfo {year} {1996}{\natexlab{a}})}\BibitemShut {NoStop}%
\bibitem [{\citenamefont {Chang}\ \emph {et~al.}(2014)\citenamefont {Chang}, \citenamefont {Sinha}, \citenamefont {Taylor},\ and\ \citenamefont {Kimble}}]{Chang14}%
  \BibitemOpen
  \bibfield  {author} {\bibinfo {author} {\bibfnamefont {D.~E.}\ \bibnamefont {Chang}}, \bibinfo {author} {\bibfnamefont {K.}~\bibnamefont {Sinha}}, \bibinfo {author} {\bibfnamefont {J.~M.}\ \bibnamefont {Taylor}},\ and\ \bibinfo {author} {\bibfnamefont {H.~J.}\ \bibnamefont {Kimble}},\ }\bibfield  {title} {\bibinfo {title} {Trapping atoms using nanoscale quantum vacuum forces},\ }\href {https://doi.org/10.1038/ncomms5343} {\bibfield  {journal} {\bibinfo  {journal} {Nature Communications}\ }\textbf {\bibinfo {volume} {5}},\ \bibinfo {pages} {4343} (\bibinfo {year} {2014})}\BibitemShut {NoStop}%
\bibitem [{\citenamefont {Fuchs}\ \emph {et~al.}(2018{\natexlab{a}})\citenamefont {Fuchs}, \citenamefont {Bennett},\ and\ \citenamefont {Buhmann}}]{Fuchs18a}%
  \BibitemOpen
  \bibfield  {author} {\bibinfo {author} {\bibfnamefont {S.}~\bibnamefont {Fuchs}}, \bibinfo {author} {\bibfnamefont {R.}~\bibnamefont {Bennett}},\ and\ \bibinfo {author} {\bibfnamefont {S.~Y.}\ \bibnamefont {Buhmann}},\ }\bibfield  {title} {\bibinfo {title} {Casimir-polder potential of a driven atom},\ }\href {https://doi.org/10.1103/PhysRevA.98.022514} {\bibfield  {journal} {\bibinfo  {journal} {Phys. Rev. A}\ }\textbf {\bibinfo {volume} {98}},\ \bibinfo {pages} {022514} (\bibinfo {year} {2018}{\natexlab{a}})}\BibitemShut {NoStop}%
\bibitem [{\citenamefont {Fuchs}\ \emph {et~al.}(2018{\natexlab{b}})\citenamefont {Fuchs}, \citenamefont {Bennett}, \citenamefont {Krems},\ and\ \citenamefont {Buhmann}}]{Fuchs18b}%
  \BibitemOpen
  \bibfield  {author} {\bibinfo {author} {\bibfnamefont {S.}~\bibnamefont {Fuchs}}, \bibinfo {author} {\bibfnamefont {R.}~\bibnamefont {Bennett}}, \bibinfo {author} {\bibfnamefont {R.~V.}\ \bibnamefont {Krems}},\ and\ \bibinfo {author} {\bibfnamefont {S.~Y.}\ \bibnamefont {Buhmann}},\ }\bibfield  {title} {\bibinfo {title} {Nonadditivity of optical and casimir-polder potentials},\ }\href {https://doi.org/10.1103/PhysRevLett.121.083603} {\bibfield  {journal} {\bibinfo  {journal} {Phys. Rev. Lett.}\ }\textbf {\bibinfo {volume} {121}},\ \bibinfo {pages} {083603} (\bibinfo {year} {2018}{\natexlab{b}})}\BibitemShut {NoStop}%
\bibitem [{\citenamefont {Sinha}\ and\ \citenamefont {Suba\c{s}\i}(2020)}]{Sinha2020PRA}%
  \BibitemOpen
  \bibfield  {author} {\bibinfo {author} {\bibfnamefont {K.}~\bibnamefont {Sinha}}\ and\ \bibinfo {author} {\bibfnamefont {Y.}~\bibnamefont {Suba\c{s}\i}},\ }\bibfield  {title} {\bibinfo {title} {Quantum brownian motion of a particle from casimir-polder interactions},\ }\href {https://doi.org/10.1103/PhysRevA.101.032507} {\bibfield  {journal} {\bibinfo  {journal} {Phys. Rev. A}\ }\textbf {\bibinfo {volume} {101}},\ \bibinfo {pages} {032507} (\bibinfo {year} {2020})}\BibitemShut {NoStop}%
\bibitem [{\citenamefont {Sinha}(2018)}]{Sinha18}%
  \BibitemOpen
  \bibfield  {author} {\bibinfo {author} {\bibfnamefont {K.}~\bibnamefont {Sinha}},\ }\bibfield  {title} {\bibinfo {title} {Repulsive vacuum-induced forces on a magnetic particle},\ }\href {https://doi.org/10.1103/PhysRevA.97.032513} {\bibfield  {journal} {\bibinfo  {journal} {Phys. Rev. A}\ }\textbf {\bibinfo {volume} {97}},\ \bibinfo {pages} {032513} (\bibinfo {year} {2018})}\BibitemShut {NoStop}%
\bibitem [{\citenamefont {Bostr{\"o}m}\ \emph {et~al.}(2012)\citenamefont {Bostr{\"o}m}, \citenamefont {Ellingsen}, \citenamefont {Brevik}, \citenamefont {Dou}, \citenamefont {Persson},\ and\ \citenamefont {Sernelius}}]{Bostron12}%
  \BibitemOpen
  \bibfield  {author} {\bibinfo {author} {\bibfnamefont {M.}~\bibnamefont {Bostr{\"o}m}}, \bibinfo {author} {\bibfnamefont {S.~{\AA}.}\ \bibnamefont {Ellingsen}}, \bibinfo {author} {\bibfnamefont {I.}~\bibnamefont {Brevik}}, \bibinfo {author} {\bibfnamefont {M.~F.}\ \bibnamefont {Dou}}, \bibinfo {author} {\bibfnamefont {C.}~\bibnamefont {Persson}},\ and\ \bibinfo {author} {\bibfnamefont {B.~E.}\ \bibnamefont {Sernelius}},\ }\bibfield  {title} {\bibinfo {title} {Casimir attractive-repulsive transition in mems},\ }\href {https://doi.org/10.1140/epjb/e2012-30794-5} {\bibfield  {journal} {\bibinfo  {journal} {The European Physical Journal B}\ }\textbf {\bibinfo {volume} {85}},\ \bibinfo {pages} {377} (\bibinfo {year} {2012})}\BibitemShut {NoStop}%
\bibitem [{\citenamefont {Buhmann}(2013{\natexlab{a}})}]{Buhmann1}%
  \BibitemOpen
  \bibfield  {author} {\bibinfo {author} {\bibfnamefont {S.}~\bibnamefont {Buhmann}},\ }\href {https://doi.org/10.1007/978-3-642-32484-0} {\emph {\bibinfo {title} {Dispersion Forces I: Macroscopic Quantum Electrodynamics and Ground-State Casimir, Casimir--Polder and van der Waals Forces}}},\ Springer Tracts in Modern Physics\ (\bibinfo  {publisher} {Springer Berlin Heidelberg},\ \bibinfo {year} {2013})\BibitemShut {NoStop}%
\bibitem [{\citenamefont {Buhmann}(2013{\natexlab{b}})}]{Buhmann2}%
  \BibitemOpen
  \bibfield  {author} {\bibinfo {author} {\bibfnamefont {S.}~\bibnamefont {Buhmann}},\ }\href {https://doi.org/10.1007/978-3-642-32466-6} {\emph {\bibinfo {title} {Dispersion Forces II: Many-Body Effects, Excited Atoms, Finite Temperature and Quantum Friction}}},\ Springer Tracts in Modern Physics\ (\bibinfo  {publisher} {Springer Berlin Heidelberg},\ \bibinfo {year} {2013})\BibitemShut {NoStop}%
\bibitem [{\citenamefont {Jackson}(1999)}]{Jackson}%
  \BibitemOpen
  \bibfield  {author} {\bibinfo {author} {\bibfnamefont {J.~D.}\ \bibnamefont {Jackson}},\ }\href {https://search.library.wisc.edu/catalog/999849741702121} {\emph {\bibinfo {title} {Classical electrodynamics}}}\ (\bibinfo  {publisher} {Third edition. New York : Wiley},\ \bibinfo {year} {1999})\BibitemShut {NoStop}%
\bibitem [{\citenamefont {Emig}\ \emph {et~al.}(2007)\citenamefont {Emig}, \citenamefont {Graham}, \citenamefont {Jaffe},\ and\ \citenamefont {Kardar}}]{Emig07}%
  \BibitemOpen
  \bibfield  {author} {\bibinfo {author} {\bibfnamefont {T.}~\bibnamefont {Emig}}, \bibinfo {author} {\bibfnamefont {N.}~\bibnamefont {Graham}}, \bibinfo {author} {\bibfnamefont {R.~L.}\ \bibnamefont {Jaffe}},\ and\ \bibinfo {author} {\bibfnamefont {M.}~\bibnamefont {Kardar}},\ }\bibfield  {title} {\bibinfo {title} {Casimir forces between arbitrary compact objects},\ }\href {https://doi.org/10.1103/PhysRevLett.99.170403} {\bibfield  {journal} {\bibinfo  {journal} {Phys. Rev. Lett.}\ }\textbf {\bibinfo {volume} {99}},\ \bibinfo {pages} {170403} (\bibinfo {year} {2007})}\BibitemShut {NoStop}%
\bibitem [{\citenamefont {Dholakia}\ and\ \citenamefont {Zem\'anek}(2010)}]{Pavel2010}%
  \BibitemOpen
  \bibfield  {author} {\bibinfo {author} {\bibfnamefont {K.}~\bibnamefont {Dholakia}}\ and\ \bibinfo {author} {\bibfnamefont {P.}~\bibnamefont {Zem\'anek}},\ }\bibfield  {title} {\bibinfo {title} {Colloquium: Gripped by light: Optical binding},\ }\href {https://doi.org/10.1103/RevModPhys.82.1767} {\bibfield  {journal} {\bibinfo  {journal} {Rev. Mod. Phys.}\ }\textbf {\bibinfo {volume} {82}},\ \bibinfo {pages} {1767} (\bibinfo {year} {2010})}\BibitemShut {NoStop}%
\bibitem [{\citenamefont {Burns}\ \emph {et~al.}(1989)\citenamefont {Burns}, \citenamefont {Fournier},\ and\ \citenamefont {Golovchenko}}]{Burns1989}%
  \BibitemOpen
  \bibfield  {author} {\bibinfo {author} {\bibfnamefont {M.~M.}\ \bibnamefont {Burns}}, \bibinfo {author} {\bibfnamefont {J.-M.}\ \bibnamefont {Fournier}},\ and\ \bibinfo {author} {\bibfnamefont {J.~A.}\ \bibnamefont {Golovchenko}},\ }\bibfield  {title} {\bibinfo {title} {Optical binding},\ }\href {https://doi.org/10.1103/PhysRevLett.63.1233} {\bibfield  {journal} {\bibinfo  {journal} {Phys. Rev. Lett.}\ }\textbf {\bibinfo {volume} {63}},\ \bibinfo {pages} {1233} (\bibinfo {year} {1989})}\BibitemShut {NoStop}%
\bibitem [{\citenamefont {Rudolph}\ \emph {et~al.}(2023)\citenamefont {Rudolph}, \citenamefont {Deli\'{c}}, \citenamefont {Hornberger},\ and\ \citenamefont {Stickler}}]{rudolph2023}%
  \BibitemOpen
  \bibfield  {author} {\bibinfo {author} {\bibfnamefont {H.}~\bibnamefont {Rudolph}}, \bibinfo {author} {\bibfnamefont {U.}~\bibnamefont {Deli\'{c}}}, \bibinfo {author} {\bibfnamefont {K.}~\bibnamefont {Hornberger}},\ and\ \bibinfo {author} {\bibfnamefont {B.~A.}\ \bibnamefont {Stickler}},\ }\href@noop {} {\bibinfo {title} {Quantum theory of non-hermitian optical binding between nanoparticles}} (\bibinfo {year} {2023}),\ \Eprint {https://arxiv.org/abs/2306.11893} {arXiv:2306.11893 [quant-ph]} \BibitemShut {NoStop}%
\bibitem [{\citenamefont {Gonzalez-Ballestero}\ \emph {et~al.}(2023)\citenamefont {Gonzalez-Ballestero}, \citenamefont {Zieli\ifmmode~\acute{n}\else \'{n}\fi{}ska}, \citenamefont {Rossi}, \citenamefont {Militaru}, \citenamefont {Frimmer}, \citenamefont {Novotny}, \citenamefont {Maurer},\ and\ \citenamefont {Romero-Isart}}]{Romero-Isart_Squeezed}%
  \BibitemOpen
  \bibfield  {author} {\bibinfo {author} {\bibfnamefont {C.}~\bibnamefont {Gonzalez-Ballestero}}, \bibinfo {author} {\bibfnamefont {J.}~\bibnamefont {Zieli\ifmmode~\acute{n}\else \'{n}\fi{}ska}}, \bibinfo {author} {\bibfnamefont {M.}~\bibnamefont {Rossi}}, \bibinfo {author} {\bibfnamefont {A.}~\bibnamefont {Militaru}}, \bibinfo {author} {\bibfnamefont {M.}~\bibnamefont {Frimmer}}, \bibinfo {author} {\bibfnamefont {L.}~\bibnamefont {Novotny}}, \bibinfo {author} {\bibfnamefont {P.}~\bibnamefont {Maurer}},\ and\ \bibinfo {author} {\bibfnamefont {O.}~\bibnamefont {Romero-Isart}},\ }\bibfield  {title} {\bibinfo {title} {Suppressing recoil heating in levitated optomechanics using squeezed light},\ }\href {https://doi.org/10.1103/PRXQuantum.4.030331} {\bibfield  {journal} {\bibinfo  {journal} {PRX Quantum}\ }\textbf {\bibinfo {volume} {4}},\ \bibinfo {pages} {030331} (\bibinfo {year} {2023})}\BibitemShut {NoStop}%
\bibitem [{\citenamefont {Boyer}(1975)}]{Boyer1975}%
  \BibitemOpen
  \bibfield  {author} {\bibinfo {author} {\bibfnamefont {T.~H.}\ \bibnamefont {Boyer}},\ }\bibfield  {title} {\bibinfo {title} {Temperature dependence of van der waals forces in classical electrodynamics with classical electromagnetic zero-point radiation},\ }\href {https://doi.org/10.1103/PhysRevA.11.1650} {\bibfield  {journal} {\bibinfo  {journal} {Phys. Rev. A}\ }\textbf {\bibinfo {volume} {11}},\ \bibinfo {pages} {1650} (\bibinfo {year} {1975})}\BibitemShut {NoStop}%
\bibitem [{\citenamefont {Milonni}\ and\ \citenamefont {Smith}(1996{\natexlab{b}})}]{Milonni1996}%
  \BibitemOpen
  \bibfield  {author} {\bibinfo {author} {\bibfnamefont {P.~W.}\ \bibnamefont {Milonni}}\ and\ \bibinfo {author} {\bibfnamefont {A.}~\bibnamefont {Smith}},\ }\bibfield  {title} {\bibinfo {title} {van der waals dispersion forces in electromagnetic fields},\ }\href {https://doi.org/10.1103/PhysRevA.53.3484} {\bibfield  {journal} {\bibinfo  {journal} {Phys. Rev. A}\ }\textbf {\bibinfo {volume} {53}},\ \bibinfo {pages} {3484} (\bibinfo {year} {1996}{\natexlab{b}})}\BibitemShut {NoStop}%
\bibitem [{\citenamefont {Passante}\ and\ \citenamefont {Spagnolo}(2007)}]{Spagnolo2007}%
  \BibitemOpen
  \bibfield  {author} {\bibinfo {author} {\bibfnamefont {R.}~\bibnamefont {Passante}}\ and\ \bibinfo {author} {\bibfnamefont {S.}~\bibnamefont {Spagnolo}},\ }\bibfield  {title} {\bibinfo {title} {Casimir-polder interatomic potential between two atoms at finite temperature and in the presence of boundary conditions},\ }\href {https://doi.org/10.1103/PhysRevA.76.042112} {\bibfield  {journal} {\bibinfo  {journal} {Phys. Rev. A}\ }\textbf {\bibinfo {volume} {76}},\ \bibinfo {pages} {042112} (\bibinfo {year} {2007})}\BibitemShut {NoStop}%
\bibitem [{\citenamefont {Hemmerich}\ \emph {et~al.}(2016)\citenamefont {Hemmerich}, \citenamefont {Bennett}, \citenamefont {Reisinger}, \citenamefont {Nimmrichter}, \citenamefont {Fiedler}, \citenamefont {Hahn}, \citenamefont {Gleiter},\ and\ \citenamefont {Buhmann}}]{PhysRevA.94.023621}%
  \BibitemOpen
  \bibfield  {author} {\bibinfo {author} {\bibfnamefont {J.~L.}\ \bibnamefont {Hemmerich}}, \bibinfo {author} {\bibfnamefont {R.}~\bibnamefont {Bennett}}, \bibinfo {author} {\bibfnamefont {T.}~\bibnamefont {Reisinger}}, \bibinfo {author} {\bibfnamefont {S.}~\bibnamefont {Nimmrichter}}, \bibinfo {author} {\bibfnamefont {J.}~\bibnamefont {Fiedler}}, \bibinfo {author} {\bibfnamefont {H.}~\bibnamefont {Hahn}}, \bibinfo {author} {\bibfnamefont {H.}~\bibnamefont {Gleiter}},\ and\ \bibinfo {author} {\bibfnamefont {S.~Y.}\ \bibnamefont {Buhmann}},\ }\bibfield  {title} {\bibinfo {title} {Impact of casimir-polder interaction on poisson-spot diffraction at a dielectric sphere},\ }\href {https://doi.org/10.1103/PhysRevA.94.023621} {\bibfield  {journal} {\bibinfo  {journal} {Phys. Rev. A}\ }\textbf {\bibinfo {volume} {94}},\ \bibinfo {pages} {023621} (\bibinfo {year} {2016})}\BibitemShut {NoStop}%
\bibitem [{\citenamefont {Aasi}\ \emph {et~al.}(2013)\citenamefont {Aasi} \emph {et~al.}}]{Aasi2013}%
  \BibitemOpen
  \bibfield  {author} {\bibinfo {author} {\bibfnamefont {J.}~\bibnamefont {Aasi}} \emph {et~al.},\ }\bibfield  {title} {\bibinfo {title} {Enhanced sensitivity of the ligo gravitational wave detector by using squeezed states of light},\ }\href {https://doi.org/10.1038/nphoton.2013.177} {\bibfield  {journal} {\bibinfo  {journal} {Nature Photonics}\ }\textbf {\bibinfo {volume} {7}},\ \bibinfo {pages} {613} (\bibinfo {year} {2013})}\BibitemShut {NoStop}%
\bibitem [{\citenamefont {Treps}\ \emph {et~al.}(2002)\citenamefont {Treps}, \citenamefont {Andersen}, \citenamefont {Buchler}, \citenamefont {Lam}, \citenamefont {Ma\^{\i}tre}, \citenamefont {Bachor},\ and\ \citenamefont {Fabre}}]{Treps2002}%
  \BibitemOpen
  \bibfield  {author} {\bibinfo {author} {\bibfnamefont {N.}~\bibnamefont {Treps}}, \bibinfo {author} {\bibfnamefont {U.}~\bibnamefont {Andersen}}, \bibinfo {author} {\bibfnamefont {B.}~\bibnamefont {Buchler}}, \bibinfo {author} {\bibfnamefont {P.~K.}\ \bibnamefont {Lam}}, \bibinfo {author} {\bibfnamefont {A.}~\bibnamefont {Ma\^{\i}tre}}, \bibinfo {author} {\bibfnamefont {H.-A.}\ \bibnamefont {Bachor}},\ and\ \bibinfo {author} {\bibfnamefont {C.}~\bibnamefont {Fabre}},\ }\bibfield  {title} {\bibinfo {title} {Surpassing the standard quantum limit for optical imaging using nonclassical multimode light},\ }\href {https://doi.org/10.1103/PhysRevLett.88.203601} {\bibfield  {journal} {\bibinfo  {journal} {Phys. Rev. Lett.}\ }\textbf {\bibinfo {volume} {88}},\ \bibinfo {pages} {203601} (\bibinfo {year} {2002})}\BibitemShut {NoStop}%
\end{thebibliography}%

\end{document}